\documentclass[useAMS,usenatbib]{aa}
\usepackage{natbib}
\usepackage{graphicx}
\usepackage{epsfig}
\usepackage{amssymb}
\usepackage{amsmath}
\usepackage{natbib}
\usepackage{times}

\usepackage[english]{babel}

\title{Subcritical transition to turbulence in accretion disc boundary layer}
\titlerunning{Cross-rolls in rotating shear flow}
\author{V. V. Zhuravlev\inst1\thanks{E-mail:
zhuravlev@sai.msu.ru}  and D. N. Razdoburdin\inst{1}}
\institute{Sternberg Astronomical Institute, Moscow M.V. Lomonosov State University, Universitetskij pr., 13, Moscow 119234, Russia}
\authorrunning{V.V. Zhuralvev and D. N. Razdoburdin}

\begin{document}

\date{
}

\label{firstpage}

\defcitealias{lesur-longaretti-2005}{LL05}
\defcitealias{zhuravlev-razdoburdin-2014}{ZR14}
\defcitealias{razdoburdin-zhuravlev-2015}{RZ15}
\abstract{

{\it Context.} 
Enhanced angular momentum transfer through the boundary layer near the surface of weakly magnetised accreting star is required in order to explain the observed 
accretion timescales in low-mass X-ray binaries, cataclysmic variables or young stars with massive protoplanetary discs.
Accretion disc boundary layer is locally represented by incompressible homogeneous and boundless flow of the cyclonic type, which is linearly stable.
Its non-linear instability at the shear rates of order of the rotational frequency remains an issue.

{\it Aims.}
We put forward a conjecture that hydrodynamical subcritical turbulence in such a flow is sustained by the non-linear feedback from 
essentially three-dimensional vortices, which are generated by quasi-two-dimensional trailing shearing spirals grown to high amplitude via the swing amplification.
We refer to those three-dimensional vortices as cross-rolls, since they are aligned in the shearwise direction in contrast to streamwise rolls generated by the anti-lift-up mechanism in rotating shear flow on the Rayleigh line.

{\it Methods.}
Transient growth of cross-rolls is studied analytically and further confronted with direct numerical simulations (DNS) of dynamics of non-linear perturbations in shearing box approximation.

{\it Results.}
A substantial decrease of transition Reynolds number $R_T$ is revealed as one changes a cubic box to a tall box. 
DNS performed in a tall box show that $R_T$ as function of shear rate accords with the line of constant maximum transient growth of cross-rolls. 
The transition in the tall box has been observed until the shear rate three times higher than the rotational frequency, when $R_T\sim 50000$.

{\it Conclusions.}
Assuming that the cross-rolls are also responsible for turbulence in the Keplerian flow, we estimate $R_T\lesssim 10^8$ in this case.
Our results imply that non-linear stability of Keplerian flow should be verified by extending turbulent solutions found in the cyclonic regime across the solid-body line rather than entering 
a quasi-Keplerian regime from the side of the Rayleigh line. The most favourable shear rate to test the existence of turbulence in the quasi-Keplerian regime may be sub-Keplerian and equal approximately to $1/2$.

}


\maketitle

\section{Introduction}

Rotating boundary layers are believed to exist in the vicinity of weakly magnetised stars accumulating the material from accretion disc.
Slowly spinning star surface causes transformation of the rotational energy of disc into thermal energy inside the boundary layer. 
Thus, the boundary layer is expected to have brightness comparable to that of the accretion disc and be the main source of radiation in relatively higher frequency band than the disc itself.
The structure of the boundary layer and the adjacent disc inner part must be determined together, see \citet{gsbk-1994}.
A number of boundary layer models have been developed for various classes of objects. 
Usually they employ the slim disc equations \citep{abramowicz-1988} in order to match consistently the disc inner part to the boundary layer. 
Among them the accreting pre-main sequence stars were studied by \citet{popham-narayan-1993}, whereas the accreting white dwarfs and neutron stars were studied by \citet{narayan-popham-1993}
and \citet{popham-sunyaev-2001}. In the latter case the formation of spreading layer is also possible in system with high enough accretion rate, as was suggested by \citet{inogamov-sunyaev-1999}.

Just as for the formation of an accretion disc, the cornerstone of boundary layer formation is the origin of effective shear viscosity providing enhanced angular momentum and mass transfer onto the star surface.
Conventionally, it is parametrised by $\alpha$, which is kinematic viscosity coefficient scaled by speed of sound and pressure radial scaleheight, see \citet{shakura-sunyaev-1988}.
This variant of viscosity prescription is similar to what is done in the disc models, where $\alpha$ is scaled by the disc thickness, see \citet{shakura-sunyaev-1973}.
However, the physical process veiled by an effective viscosity as well as magnitude of $\alpha$ remain a matter of debate, probably, to a greater extent in the case of boundary layer rather than in the case of accretion disc.
Turbulence is known to be natural solution to this issue, but its simple (supercritical) variant is discarded by centrifugal stability of the rotating shear flows, which represent both accretion discs and boundary layers. 
A `magic wand' working in hot magnetised discs, where angular velocity $\Omega$ decreases with distance to the rotation axis $r$, is a magnetorotational instability. 
However, it does not work if $d\Omega^2/dr > 0$, which is the case for the boundary layers. Such flows are locally linearly stable even in the presence of the magnetic field.
In this context, an alternative mechanism of angular momentum transfer have been proposed by \citet{rafikov-2013} and more recently by \citet{philippov-2016}.
It was shown that global sonic instability of the flow associated with the presence of the star surface excites global non-axisymmetric acoustic modes, which are responsible for effective viscosity even in the absence of 
turbulence. 

At the same time, the possibility that astrophysical boundary layers acquire the effective viscosity through the turbulence is not ruled out, 
since flows with shear rates including that of order of $|\Omega|$ may be locally non-linearly unstable.
Along with hydrodynamical non-linear stability of homogeneous Keplerian shear flow, which locally represents the accretion and protoplanetary discs, 
this is one of the major unresolved problems in astrophysical fluid dynamics. Below we consider both problems jointly.
In order to be non-linearly unstable, such flows at first require a candidate for transiently growing perturbations, which could be amplified by background shear by orders of magnitude for Reynolds number $R$ typical in astrophysical situation; at second, in phase of high amplitude, those perturbations have to give a positive non-linear feedback to the basin of small perturbations subject to transient growth. 
Two of these conditions together introduce what is usually called a bypass scenario for transition to turbulence. 

The bypass scenario is best developed in application to plane shear flows and, in particular, to the Couette
flow, see \citet{waleffe-1995} and \citet{waleffe-1997}, who proposed a non-linear mechanism sustaining the turbulence. This mechanism is called self-sustaining process (SSP) and is sketched, for example in Fig. 1 by \citet{waleffe-1997}.
The transient growth of initially small streamwise rolls produces high amplitude streamwise streaks, the process usually called lift-up effect after the work by \citet{ellingsen-palm-1975}.
As streamwise streaks modify the background velocity yielding inflexion point in spanwise direction, the flow becomes linearly unstable with respect to non-axisymmetric modes, which in turn regenerate streamwise rolls via non-linear interactions with each other. 

However, the situation becomes less understood in centrifugally stable rotating shear flows.
Let us define rotation number, which characterises a dynamical contribution of rotation to shear,
$
R_\Omega = -2/q,
$
where $q = -(r/\Omega) d \Omega / d r$ is the local dimensionless shear rate.
Steady non-linear self-sustaining solutions obtained in the non-rotating case, for which $R_\Omega=0$, 
can be non-linearly continued into both the centrifugally unstable regime, $-1<R_\Omega<0$ ($q>2$), and the cyclonic regime with high shear rate, $R_\Omega\ll 1$ ($-q\gg 1$), see \citet{rincon-2007}.
Nevertheless, this operation cannot be extended neither until the centrifugally stable anti-cyclonic (or quasi-Keplerian) regime with $R_\Omega \leq -1$ ($0<q<2$), nor far into the cyclonic regime with $R_\Omega\lesssim 1$ ($q\lesssim -1$). 
Presumably, this takes place because the lift-up mechanism ceases to work in both cases. 
On the other hand, direct numerical simulations (DNS) allow to obtain turbulent solutions until, at least, $R_\Omega \approx 0.3$, see \citet{lesur-longaretti-2005} (\citetalias{lesur-longaretti-2005} hereafter). 
It is interesting to note that such a value of rotation number is far beyond the interval of small positive $R_\Omega$ corresponding to a substantial transient growth of streamwise rolls, see Section \ref{TG_rolls} of this paper.
Another notable fact is that the quasi-Keplerian regime demonstrates a strong asymmetry in comparison with the cyclonic regime. Namely, DNS show
dramatic stabilisation of quasi-Keplerian shear, as one goes away from the Rayleigh line, $R_\Omega=-1$, to smaller rotation numbers, cf. Fig. 4 and Fig. 7 of \citetalias{lesur-longaretti-2005}.
See also the earlier DNS of quasi-Keplerian regime made by \citet{hawley-1999}.
We have already noted that when $R_\Omega=-1$ exactly, the lift-up mechanism does not work, however, there is an inverse process of small streamwise streaks growing into high amplitude streamwise rolls, which
is called anti-lift-up effect, see Section 2.3 in \citet{rincon-2008} for more detail. The regime $R_\Omega=-1$ is neutrally stable with respect to local modal perturbations, i.e. the growth of streaks 
can be regarded as epicyclic oscillations with an infinite period. Because finite amplitude streamwise rolls of the anti-lift-up are quite different from finite amplitude streamwise streaks of the lift-up, 
\citet{rincon-2007} put forward several arguments that the turbulent solution should be represented by a bypass scenario another than SSP. In any case, since a bypass transition in the regime $R_\Omega=-1$ should involve the anti-lift-up mechanism, 
it looks natural that \citetalias{lesur-longaretti-2005} observed a sharp increase of the transition Reynolds number $R_T$ as one goes to $R_\Omega < -1$, in contrast to the situation with the cyclonic regime. 
Indeed, the streaks growth is strongly suppressed at $R_\Omega < -1$, as it turns into epicyclic oscillations with nearly rotational frequency (see also Section \ref{TG_streakes} of this paper).
So, \citetalias{lesur-longaretti-2005} observed a decay of initial turbulence in the ranges $-\infty < R_\Omega < -1.03$ and $0.3 < R_\Omega < \infty$ for $R\lesssim 10^5$ within a local model of the shearing box with
shearing boundary conditions. A bypass scenario proposed in this work naturally resolves the issues mentioned above.

There is a number of studies that examined properties of centrifugally stable flows either experimentally, or employing DNS in the framework of Taylor-Couette problem, i.e. the flow between the coaxial corotating cylinders, 
see Section 6 of the review by \citet{grossmann-2016}. Among the recent ones are \citet{burin-2012} who observed the turbulisation of the cyclonic flow at $R_T\sim 10^5$ 
in the case of $R_\Omega\approx 0.8$ and \citet{ostilla-2016} who considered the non-stationary dynamics in the same regime numerically excluding the influence of axial boundaries, which is unavoidable in the laboratory setup. 
At the same time, the quasi-Keplerian regime was investigated in the experiment by \citet{schartman-2012} subsequently continued by \citet{edlund-ji-2014} and numerically by \citet{ostilla-2014} subsequently continued by \citet{avila-2017}. 
All those studies demonstrated an outstanding stability of quasi-Keplerian flow with respect to finite amplitude perturbations from $R\sim 10^5$ up to higher than $R\sim 10^6$. 
Yet, this result should be accepted with a caution, because the presence of walls may prevent the onset of self-sustaining solutions inherent for anti-cyclonic subcritical regime as was argued by \citet{rincon-2007}.
Additionally, let us emphasize that most of attempts to detect turbulence have been made at super-Keplerian shear rates, $3/2<q<2$.
For example, \citet{edlund-ji-2014} and \citet{avila-2017} used $q=1.8$ and $q=5/3$, respectively\footnote{Actually, these are the radially averaged shear rates.}. Indeed, there is a tendency to consider the shear rates starting from $q=2$, which stems from the established opinion that the closer quasi-Keplerian shear to the Rayleigh line, $R_\Omega=-1$ ($q=2$), and the farther it from the solid-body line, $R_\Omega\to-\infty$ ($q=0$), the less stable it becomes.
However, an accompanying purpose of this work is to show that the most appropriate shear rate to test the transition to turbulence in the quasi-Keplerian regime may be found in the vicinity of $q=1/2$, see Section \ref{prediction}.

Two decades ago it was established that shear flow turbulence is a result of subtle interplay between non-normal linear dynamics of perturbations and their 
non-linear interaction, see e.g. the review by \citet{grossman-2000}. Although the subcritical transition is in essence a loss of non-linear stability,
the phase of linear, non-modal transient growth is of much importance in general design of shear flow turbulence. Among others, this was elucidated by \citet{henningson-1996}. 
It was argued that
(i) transient growth as linear process is the only source of energy for subcritical turbulence, 
(ii) transient growth must be large, i.e. the linear growth factor must be much greater than unity, for the transition to occur:
for the particular examples, the plane Couette and Poiseille flows may become turbulent at the Reynolds numbers as low as, respectively, $R_T\simeq 350$ and $1000$, when
according to \citet{butler-farrell-1992} and \citet{reddy-henningson-1993} the maximum growth factor of streamwise rolls is, respectively, $\simeq 150$ and $200$.
Also, a number of low-dimensional models of plane shear flow employing various representations of the non-linear dynamics
demonstrated qualitatively similar behaviour during the transition: the substantial non-modal growth followed by the 'bootstrapping' effect \citep{trefethen-1993}, see \citet{baggett-trefethen-1997}. 
These authors explained threshold exponents for transition to turbulence in each case using simple arguments based on the secondary
linear instability of the transient perturbation attained high amplitude. Thus, it was shown that 
transient dynamics can always be distinguished in the whole loop of the sustenance of turbulence, see caption for Fig. 3 by \citet{baggett-trefethen-1997}.

The reasoning just mentioned makes the transient growth a pivotal process responsible for generation of the specific high-amplitude inputs for the sequential non-linear feedback.
Such high-amplitude inputs are the streamwise streaks in the case of plane shear flows, but what are they in the case of centrifugally stable homogeneous and boundless rotating 
shear flow? To answer this question in this paper, we (i) consider three-dimensional (3D) perturbations, which are able to exhibit large transient growth, 
(ii) perform a critical test, which indicates that high-amplitude perturbations generated during this transient growth have to do with the sustenance of turbulence. 
The issues (i) and (ii) are tackled solving, respectively, the linear problem (\ref{sys1}-\ref{sys4}) and the more general non-linear problem (\ref{continuity}-\ref{eos}).

It is known that shear flow allows for the existence of the so called Kelvin modes, or alternatively, shearing vortices exhibiting the transient growth via the swing amplification\footnote{In hydrodynamical literature it is also called the Orr mechanism.}, see \citet{chagelishvili-2003}.
The shearing vortices are spanwise invariant nearly streamwise perturbations contracted by a background motion. In rotating flow they take the form of spirals, whereas
the instant of swing corresponds to transformation of the leading spirals into trailing spirals or vice versa, see e.g. \citet{razdoburdin-zhuravlev-2015}. Conservation of a spanwise vorticity perturbation is the physical reason 
for the growth of their velocity and pressure perturbations. The swing amplification is a two-dimensional (2D) process, which takes place independently of whether the shear is plane or rotating: the Coriolis force does not 
affect the growth of shearing vortices in contrast to previously discussed rolls and streaks. However, shearing vortices grow considerably weaker rather than rolls and streaks, since they are much more suppressed 
by viscosity. The swing amplification gives the kinetic energy growth by a factor $\propto R^{2/3}$, rather than $\propto R^2$ as in the case of lift- and anti-lift-up. 
\citet{mukhopadhyay-2006} estimated that $R\sim 10^6$ is required to obtain the growth factor of shearing spirals comparable to the growth factor of rolls necessary to activate the SSP in the plane Couette flow.
Remarkably, this claim is valid for both the plane and the rotating (no care of which cyclonicity) flows and does not change with the shear rate, since it can be shown that the maximum growth factor of 
shearing vortices $G_{\max}\approx R^{2/3}\exp{(-2/3)}$ provided that the time is measured in the inversed shear rate rather than in the inversed rotational frequency, cf. equation (83) by \citet{afshordi-2005} and
equation (9) by \citet{mukhopadhyay-2005}.  
Thus, the role of shearing vortices is negligible either at the Rayleigh line, or in absence of rotation, but as soon as $R_\Omega$ is either sufficiently smaller than $-1$, or sufficiently larger than $0$, 
the growth of streaks and rolls vanishes and the shearing vortices become the kind of perturbations exhibiting the largest transient growth in the flow. This picture was confirmed for a variety of models by \citet{ioannou-kakouris-2001}, \citet{mukhopadhyay-2005}, \cite{yecko-2004} and more recently by \citet{maretzke-2014}, \citet{zhuravlev-razdoburdin-2014} and \cite{razdoburdin-zhuravlev-2017} who carried out studies of transient dynamics of linear perturbations 
in quasi-Keplerian regime using rigorous optimisation methods, which allow to obtain the optimal perturbations exhibiting the largest possible transient growth.

For very high Reynolds numbers typical in astrophysical boundary layers and the adjacent accretion discs the growth of shearing spirals becomes quite significant. However, its role in the transition to turbulence remains unclear. 
On the one hand, it seems plausible that they are involved in 2D models of subcritical turbulence in shear flows simulated by \citet{umurhan-regev-2004}, \citet{johnson-gammie-2005b} and \citet{horton-2010}. 
Furthermore, \citet{lithwick-2007} suggested a 2D variant of positive non-linear feedback sustaining the growth of shearing spirals.
He showed that the coupling between shearing spirals at the instant of swing and small-amplitude axisymmetric vortices generates a new shearing spiral capable of transient growth.
On the other hand, \citet{lithwick-2009} revealed that the streamwise scale of shearing spiral must exceed the disc scaleheight to make this process working in 3D model, otherwise
the shearing spiral is destroyed by resonant interaction with 3D axisymmetric vortex; see also the simulations by \citet{shen-2006}.
Consequently, it seems reasonable that shearing spirals are only relevant to 2D turbulence, which may occur at scales above the accretion disc scaleheight: moreover,
\citet{razdoburdin-zhuravlev-2017} have recently confirmed that shearing spirals keep the ability for significant transient growth even when their streamwise scale is comparable to disc global radial scale.
Also, the lack of the positive 3D feedback from shearing spirals at phase of swing is indirectly confirmed by DNS of \citetalias{lesur-longaretti-2005}, which show that $R_T$ substantially depends on the
shear rate for $R_\Omega\lesssim 1$ in the cyclonic case. In turn, this means that $G_{max}$ strongly varies along the curve of $R_T(q)$. Hence, the positive non-linear feedback would come into play at
different amplitude of working perturbation $\propto \sqrt{G_{max}}$ for different shear rates, which 
would require further clarification\footnote{Especially, bearing in mind a result of \citet{meseguer-2002} who revealed that subcritical transition observed in counter-rotating flow between the cylinders occurs along the line of constant $G_{max}$ on the plane of the Reynolds numbers corresponding to inner and outer cylinders, see Fig. 2 therein.}.
Additionally, \citet{maretzke-2014} revealed a light difference between the optimal perturbations in cyclonic and quasi-Keplerian regimes of the flow between the cylinders: 
in the cyclonic regime the optimals are not shearing spirals but their slightly modified counterparts with a weak spanwise dependence.
Coincidently or not, the non-linear instability of the cyclonic flow between cylinders is already observed, rather than that of the quasi-Keplerian flow. 
Summarizing, we suppose that the shearing spirals considered as spanwise invariant perturbations seem unlikely to be an ingredient of the non-linear feedback in the bypass scenario of subcritical transition to 3D turbulence on 
scales small compared to the disc scaleheight.

In this work we consider analytically the linear local dynamics of quasi-2D shearing spirals in the rotating, homogeneous, boundless and centrifugally stable shear flow. 
By quasi-2D shearing spiral we mean a spatial Fourier harmonics (SFH) of small perturbation, which initially has a zero spanwise velocity perturbation, but a small non-zero spanwise wavenumber.
Such perturbations are not optimal, i.e. they do not attain the largest possible growth factor measured at the instant of swing. Instead, during the swing amplification they generate essentially 3D vortices of a new type, 
which we refer to as cross-rolls, since those are aligned in the shearwise direction in contrast to ordinary streamwise rolls generated via the anti-lift-up mechanism on the Rayleigh line, see Fig. \ref{fig_4_5}.
We show that while the 'planar' eddies constructed of shearwise and streamwise velocity perturbations decay as the shearing spiral is wound back by a background flow, the cross-rolls conserve their amplitude and
fade due to viscous dissipation only. The growth factor of cross-rolls with respect to the growth factor of shearing spiral at the instant of swing is less than unity and decreases when the shear rate
alters from $q=-\infty$ up to $q=2$. Thus, the cross-rolls amplitude is larger in the cyclonic regime, but smaller in the quasi-Keplerian regime, and vanishes both at the solid-body line and at the Rayleigh line.
Note that the growth factor of cross-rolls can be arbitrarily high provided that the initial quasi-2D shearing spiral is sufficiently tightly wound.
We suggest that the non-linear instability and corresponding self-sustained turbulent solutions in the regime sufficiently far from both the centrifugally unstable interval and the non-rotating case are provided by the positive non-linear feedback associated with either a secondary linear instability of the flow containing the finite amplitude cross-rolls, or the interaction of cross-rolls with each other. Either of these two processes may generate new small quasi-2D shearing spirals 
capable of substantial swing amplification. Evidence for that comes from our simulations of subcritical transition to turbulence in the cyclonic regime. It turns out that in the shearing box model with periodic
boundary conditions the transition is facilitated in the box stretched along the rotation axis, which indicates that perturbations with small but non-zero spanwise wavenumber are involved in that process. 
Indeed, the numerical domain of finite size does not affect the 2D shearing spirals while it works as an ideal high-pass filter for harmonics with non-zero spanwise wavenumbers: 
it completely cuts off wavelengths larger than spanwise size of the numerical domain, whereas the first wavelength smaller than this size and
satisfying the periodic boundary conditions is the best resolved by the numerical scheme. 
The inherent 'flaw' of shearing box simulations becomes the useful tool in our case: the cross-rolls with streamwise size comparable to 
the box streamwise size, so the least susceptible to viscous damping, are forbidden in a cubic box but allowed in a tall box.
This way, in the tall box we manage to find the transition up to $q=-3$ ($R_\Omega\approx 0.67$). 
Furthermore, we find that $R_T$ obtained in the tall box simulations corresponds to approximately constant maximum growth factor of cross-rolls in the range $q\sim -35\div-3$.   
Supposing that the transition to self-sustained turbulence in quasi-Keplerian regime corresponds to the same threshold growth factor of cross-rolls, we predict the transition Reynolds number for $0<q<2$, see Fig. \ref{fig_8}, 
which is $R_T\lesssim 10^8$ for the Keplerian shear rate. Note that it is consistent with the results of \citetalias{lesur-longaretti-2005} concerning the asymmetry of $R_T(q)$ in the
vicinity of $R_\Omega=0$ and $R_\Omega=-1$. Besides, it seems unrealistic to obtain the transition in quasi-Keplerian regime considering the super-Keplerian shear rates. 
If our suggestion is true, the strategy for search of hydrodynamical turbulence in Keplerian flow should be changed: one should extend the turbulent
solution from the cyclonic regime with high negative shear rate to quasi-Keplerian regime across the solid-body line, $R_\Omega=+\infty\to R_\Omega=-\infty$, rather than to start from a turbulent solution obtained on the Rayleigh line, which has a different origin and properties. We predict that the most favourable value of the shear rate to test the existence of hydrodynamical turbulence in quasi-Keplerian regime is $q=1/2$. 

Note that in the case of plane shear flow, $R_\Omega=0$, SFH with arbitrary non-zero spanwise wavenumber is described by an exact analytical solution obtained by \citet{chagelishvili-2016}. 
In the limit of zero spanwise wavenumber quasi-2D shearing spirals turn into SFH, which are usually called shearing spirals. From now on, we omit prefix 'quasi-2D', so that by shearing spirals we mean all SFH with initially planar velocity perturbation, which may have small non-zero spanwise wavenumber.

\section{Linear perturbations in rotating shear flow}
\label{lin_perts}

Local vortical 3D perturbations in viscous homogeneous rotating shear flow obey the following equations:
\begin{equation}
\label{sys1}
\left ( \frac{\partial}{\partial t} - q\Omega_0 x\frac{\partial}{\partial y} \right ) u_x - 2\Omega_0 u_y =
-\frac{1}{\rho_0} \frac{\partial p}{\partial x} + \nu \nabla^2 u_x,
\end{equation}
\begin{equation}
\label{sys2}
\left ( \frac{\partial}{\partial t} - q\Omega_0 x\frac{\partial}{\partial y} \right ) u_y + 
(2 - q)\Omega_0 u_x =
-\frac{1}{\rho_0} \frac{\partial p}{\partial y} + \nu \nabla^2 u_y,
\end{equation}
\begin{equation}
\label{sys3}
\left ( \frac{\partial}{\partial t} - q\Omega_0 x\frac{\partial}{\partial y} \right ) u_z =
-\frac{1}{\rho_0} \frac{\partial p}{\partial z} + \nu \nabla^2 u_z,
\end{equation}
\begin{equation}
\label{sys4}
\frac{\partial u_x}{\partial x} + \frac{\partial u_y}{\partial y} + \frac{\partial u_z}{\partial z} = 0,
\end{equation}
where $u_x,u_y$ and $u_z$ are the Eulerian perturbations of velocity components, $\rho_0$ is a constant background density and $p$ is the Eulerian perturbation of pressure.
By definition, 
$$
\nabla^2 \equiv \frac{\partial^2}{\partial^2 x} + \frac{\partial^2}{\partial^2 y} + \frac{\partial^2}{\partial^2 z}
$$
and kinematic viscosity $\nu$ is assumed to be a constant. 
Variables $x$, $y$ and $z$ are Cartesian coordinates, which locally correspond to radial, azimuthal and axial directions in boundary layer (or disc), respectively, whereas
$\Omega_0$ is angular velocity of fluid rotation at $x=0$ corresponding to some radial distance $r_0\gg x$. Also, $q$ is a constant shear rate that defines 
the background azimuthal velocity as $v_y=-q\Omega_0 x$. Equations (\ref{sys1}-\ref{sys4}) is vertically unstratified linearised version of small shearing box equations, see A.3 \citet{umurhan-regev-2004}.
Note that throughout the text the variables $x,y,z$ are alternatively referred to as, respectively, shearwise, streamwise and spanwise coordinates (or directions).
The shear rate is positive (negative) in the case of anticyclonic (cyclonic) flow. The anticyclonic case is usually divided into Rayleigh-stable regime $0<q\leq 2$, 
also referred to as quasi-Keplerian regime with particular $q=3/2$ corresponding to the Keplerian flow, 
and Rayleigh-unstable regime $q>2$ subject to centrifugal linear instability. The solid-body line and the Rayleigh line are represented, respectively, by $q=0$ and $q=2$, whereas $q\to \pm\infty$ are the limits corresponding to plane shear flow. 
In what follows, we are interested in the whole range $-\infty<q<2$.

\subsection{Dimensionless equations for shearing harmonics}

We consider a boundless shear, which allows us to look for the partial solutions of equations (\ref{sys1}-\ref{sys4}) in the form of SFH,
\begin{equation}
\label{f}
f = \hat f (k_x,k_y,k_z,t^\prime) \exp ({\rm i} k_x x^\prime + {\rm i} k_y y^\prime + {\rm i} k_z z^\prime),
\end{equation}
where $f$ is any of perturbation quantities and $\hat f$ is its Fourier amplitude specified by the dimensionless wavenumbers $(k_x, k_y, k_z)$.
By equation (\ref{f}) it is implied that problem is considered with respect to the dimensionless comoving Cartesian coordinates:
\begin{equation}
\label{dim_coords}
x^\prime =  x/L,\, y^\prime = (y+q\Omega_0 xt)/L,\, z^\prime = z/L, t^\prime=\Omega_0 t,
\end{equation}
where $L$ is an auxiliary distance. Correspondingly, $k_x$, $k_y$ and $k_z$ are expressed in units of $L^{-1}$. 
Let us assume from now on that $k_y>0$ and $L$ is chosen in such a way that $k_y=1$.
We omit the primes everywhere below and note that with respect to the original coordinates SFH acquires a varying radial wavenumber $\tilde k_x \equiv k_x + q t$. 
We arrive at the following dimensionless equations for SFH

\begin{equation}
\label{orig_eq_1}
\frac{d\hat u_x}{dt} = 2\hat u_y - {\rm i}\tilde k_x \hat W - R^{-1} (k^2 + k_z^2) \hat u_x,
\end{equation}
\begin{equation}
\label{orig_eq_2}
\frac{d\hat u_y}{dt} = -(2-q)\hat u_x - {\rm i} \hat W - R^{-1} (k^2 + k_z^2) \hat u_y,
\end{equation}
\begin{equation}
\label{orig_eq_3}
\frac{d\hat u_z}{dt} = -{\rm i} k_z \hat W - R^{-1} (k^2 + k_z^2) \hat u_z,
\end{equation}
\begin{equation}
\label{orig_eq_4}
\tilde k_x \hat u_x + \hat u_y + k_z \hat u_z = 0,
\end{equation}
where velocities are expressed in units of $\Omega_0 L$ and pressure normalised by a constant background density, $W\equiv p/\rho_0$, is expressed in units of $L^2\Omega_0^2$.
The non-zero kinematic viscosity yields in a finite value of the Reynolds number
\begin{equation}
\label{Re_def}
R \equiv \frac{\Omega_0 L^2}{\nu}
\end{equation}
and $
k^2\equiv 1 + \tilde k_x^2.
$

Note that the normalisation used above is somewhat different from the accepted one, see e.g. \citetalias{lesur-longaretti-2005} and \citet{mukhopadhyay-2005}. Specifically, the time is measured
in units of $\Omega_0^{-1}$ rather than in units of $(q\Omega_0)^{-1}$. We choose the former variant, since it allows us to cross continuously the solid-body line.

Next, as soon as we are dealing with vortical dynamics, it is convenient to operate with equations for $\hat u_x$ and a shearwise component of vorticity SFH defined as $\mbox{\boldmath$\hat\omega$}\equiv {\bf k}\times \hat {\bf u}$:
\begin{equation}
\label{vort_x}
\hat\omega_x = \hat u_z - k_z \hat u_y. 
\end{equation}

Equations (\ref{orig_eq_1}-\ref{orig_eq_4}) yield
\begin{equation}
\label{rot_eq_1}
\frac{d\hat u_x}{dt} = -2\frac{k_z}{k^2+k_z^2} \hat\omega_x - 2q\frac{\tilde k_x}{k^2+k_z^2} \hat u_x - R^{-1} (k^2 + k_z^2) \hat u_x,
\end{equation}
\begin{equation}
\label{rot_eq_2}
\frac{d\hat \omega_x}{dt} = (2-q) k_z \hat u_x - R^{-1} (k^2 + k_z^2) \hat \omega_x.
\end{equation}

\section{Production of 3D vortices by shearing spirals: inviscid dynamics}

Let us assume throughout this Section that $R\to\infty$.

\subsection{Swing amplification}
\label{sw_ampl}

Here we revisit the well-known mechanism of swing amplification of vortices which takes place in $xy$-plane.
Namely, in the case $\hat u_z=0$, $k_z=0$ the problem (\ref{rot_eq_1}-\ref{rot_eq_2}) becomes one-dimensional and the evolution of the corresponding 2D SFH is described by a single equation
\begin{equation}
\label{2D}
\frac{d \hat u_x}{dt}=-2q\frac{\tilde k_x}{k^2} \hat u_x,
\end{equation}
provided that the velocity is divergence-free, 
\begin{equation}
\label{div_free}
\hat u_y = -\tilde k_x \hat u_x.
\end{equation}

The solution of equation (\ref{2D}) is
\begin{equation}
\label{u_x_analyt}
\hat u_x = \frac{k_0}{k^2},
\end{equation}
where the normalisation $k_0\equiv k(t=0)$ is obtained from the condition of doubled kinetic energy density of 2D SFH equals to unity:
$$
(\hat u_x^2 + \hat u_y^2)|_{t=0} = 1.
$$

Accordingly, we have
\begin{equation}
\label{u_y_analyt}
\hat u_y = -\tilde k_x\, \frac{k_0}{k^2}.
\end{equation}

The perturbation of pressure behaves in the following way
\begin{equation}
\label{W_analyt}
\hat W = 2{\rm i} \frac{k_0}{k^4} (k^2 - q).
\end{equation}

The solution (\ref{u_x_analyt}-\ref{u_y_analyt}) conserves spanwise component of the vorticity (see \citet{chagelishvili-2003}):
\begin{equation}
\label{om_z_2D}
\hat \omega_z = \tilde k_x \hat u_y - \hat u_x = -k_0,
\end{equation}
which results in the transient growth of 2D SFH in the case of leading spirals, $k_x < 0$, for the quasi-Keplerian regime, $q>0$, and in the case of trailing spirals, $k_x>0$, for the cyclonic regime, $q<0$. 
Indeed, at the instant of swing, $t_s\equiv - k_x/q$, the shearwise wavenumber vanishes, $\tilde k_x=0$, and their growth factor defined as the ratio
of current energy density of 2D SFH to its initial energy density reads
\begin{equation}
\label{TG_2D}
g_s \equiv g|_{t=t_s} \equiv (\hat u_x^2 + \hat u_y^2)|_{t=t_s} = \frac{k_0^2}{k^2(t=t_s)} = k_0^2.
\end{equation}
For $ |k_x|\gg 1$ $g_s \approx k_x^2$ attaining an arbitrary high value for sufficiently wound spirals. The maximum possible $g_s$ is limited by damping due to microscopic viscosity of the fluid and
will be evaluated in Section \ref{sect_visc}.

\subsection{Dynamics of SFH with non-zero spanwise wavenumber}
\label{vert_dyn}

Let us consider SFH with small but non-zero $|k_z| \ll 1$\footnote{Note that a rigorous restriction on the value of $k_z$ including the possible smallness of $q$ will be made after we have the corresponding modified solution at hand, see Section \ref{validity}.}. Without loss of generality we assume that $k_z>0$ from now on.
We also assume that initially SFH has a planar velocity field, i.e. $\hat u_z(t=0)=0$. Combining (\ref{orig_eq_4}) and (\ref{vort_x}) we formulate the initial condition in the form:
\begin{equation}
\label{init_cond}
\hat u_x = k_0^{-1}, \quad \hat \omega_x = k_z k_x k_0^{-1}.
\end{equation}
It can be seen that equation (\ref{rot_eq_2}) together with the initial condition (\ref{init_cond}) implies that $\hat\omega_x \sim k_z$, 
which makes the first term in the RHS of equation (\ref{rot_eq_1}) to be as small as $\sim O(k_z^2)$ comparing to the second term therein.
Consequently, in order to obtain the solution for 3D SFH in the leading order in $k_z\ll 1$, we drop the first term in the RHS of equation (\ref{rot_eq_1}) going back to 2D solution for $\hat u_x$ given by equation (\ref{2D}).
Thus, for SFH weakly depending on spanwise direction $\hat u_x(t)$ is approximately given by equation (\ref{u_x_analyt}). 

Next, combining (\ref{orig_eq_4}) and (\ref{vort_x}) we get an exact expression for $\hat u_z$ in terms of $\hat u_x$ and $\hat\omega_x$:
\begin{equation}
\label{u_z}
\hat u_z = \frac{\hat\omega_x - k_z\tilde k_x \hat u_x}{1 + k_z^2}.
\end{equation}
Thereby, the term $k_z \hat u_z$ in the continuity equation (\ref{orig_eq_4}) must be as small as $\sim O(k_z^2)$ in comparison with the other two terms there.
Thus, we assume that the relation (\ref{div_free}) holds on for 3D SFH and $\hat u_y$ is approximately given by its 2D expression, (\ref{u_y_analyt}). 
Of course, this implies that pressure also approximately obeys the 2D solution (\ref{W_analyt}).

As we see, spanwise dynamics of 3D SFH is separated from the swing amplification dynamics described in Section \ref{sw_ampl} and $\hat \omega_x$ is obtained from equation (\ref{rot_eq_2}) by integrating the expression (\ref{u_x_analyt}) 
over the time,
\begin{equation}
\label{analyt_om_x}
\hat \omega_x=k_z k_0 \left[\frac{2-q}{q}\left(\arctan \tilde k_x-\arctan k_x\right)+\frac{k_x}{k_0^2}\right].
\end{equation}
As $t\to\infty$ and, respectively, $|\tilde k_x|\to\infty$, the shearwise component of the vorticity perturbation tends to a constant non-zero value. 
We refer to this asymptote of SFH as a 'plateau' of SFH. 
We conclude that along with the existing 'planar' eddies, a set of another vortices associated with this non-zero limit of $\omega_x$ is generated as a 'byproduct' of swing amplification. The latter will be referred to as 'cross-rolls' in this work. The cross-rolls are directed across the stream in contrast to the streamwise rolls produced by the anti-lift-up mechanism in the flow on the Rayleigh line, see Fig. \ref{fig_4_5} and the details in Appendix \ref{app_represent} below. An important feature of the cross-rolls is that the spanwise velocity perturbation gains a non-zero value at $t\to\infty$, whereas the velocity perturbation components associated with the planar eddies 
vanish like $\hat u_x \sim O(t^{-2})$ and $\hat u_y \sim O(t^{-1})$.
Equations (\ref{u_x_analyt}) and (\ref{analyt_om_x}) together with the relation (\ref{u_z}) taken in the leading order in $k_z\ll 1$ yield\footnote{Alternatively, 
$\hat u_z$ can be obtained by integrating equation (\ref{orig_eq_3}) with the RHS substituted from (\ref{W_analyt}).}
\begin{equation}
\label{u_z_analyt}
\begin{aligned}
\hat u_z= k_z k_0 \left[ \frac{2-q}{q}\left(\arctan \tilde k_x-\arctan k_x\right) + \frac{k_x}{k_0^2}-\frac{\tilde k_x}{k^2}\right].
\end{aligned}
\end{equation}
Thus, equation (\ref{u_z_analyt}) leads to the following asymptote of $\hat u_z$ at $t\to\infty$
\begin{equation}
\label{u_z_analyt_lim}
\hat u_{z} \to {\rm sgn}(q) \pi k_z |k_x| \frac{2-q}{q}
\end{equation}
As soon as the shearing spiral is wound up by the shear after the instant of swing, the ratio $|\hat u_z/\hat u_y|$ increases, which implies that the cross-section of cross-rolls becomes elongated in spanwise direction. At the same time, cross-rolls shrink along their axes, as far as the shearwise wavenumber grows up. This, in turn, makes a streamwise component of the vorticity perturbation, $\omega_y$, to grow linearly with time. 
Since
\begin{equation}
\label{analyt_om_y}
\hat \omega_y = k_z \hat u_x - \tilde k_x \hat u_z,
\end{equation}
\begin{equation}
\label{omega_y_lim}
\hat \omega_y \to - {\rm sgn}(q) \pi k_z |k_x| (2-q)\, t.
\end{equation}

Equation (\ref{u_z_analyt_lim}) demonstrates the basic features of cross-rolls production. Clearly, their amplitude is proportional to pressure perturbation spanwise gradient at the instant of swing, which is $\propto k_z |k_x|$, cf. equation (\ref{W_analyt}) at $\tilde k_x=0$. Also, it decreases as the shear rate goes from $q=-\infty$ to $q=2$ and vanishes exactly at $q=2$. Those features will be illustrated below in Fig. \ref{fig_1}, where the growth factors of shearing spirals with non-zero spanwise number will be 
plotted for different shear rates.

\subsection{Restriction on value of spanwise wavenumber}
\label{validity}

Now we are in a step away to estimate the maximum value of $k_z=k_{z\,\max}$ which restricts the validity of the solution for 3D SFH obtained above in the leading order in $k_z\ll 1$.
At sufficiently large $k_z$, the spanwise dynamics cannot be separated from the swing amplification anymore, starting to destruct the shearing spiral in $xy$-plane.
Moreover, comparing the first term in RHS of equation (\ref{rot_eq_1}) to the second term therein, we find that these terms always become of the same order at a sufficiently large time 
for any small value of $k_z$ provided that the first term contains $\hat\omega_x$ given by equation (\ref{analyt_om_x}), while the second term contains $\hat u_x$ given by equation (\ref{u_x_analyt}). 
This takes place since $\hat u_x$ decreases as $O(t^{-2})$ long after the instant of swing. We must consequently provide a restriction on $k_z$ estimating a higher order correction to the
solution given in Sections \ref{sw_ampl} and \ref{vert_dyn} at the selected period of time. 

For that, let us assume that with the account of the first term in the RHS of equation (\ref{rot_eq_1}) the 2D solution (\ref{u_x_analyt}) acquires a small correction $k_z^2 \hat u_x^\prime$.
Substituting 
$$
\hat u_x = \frac{k_0}{k^2} + k_z^2 \hat u_x^\prime
$$
into equation (\ref{rot_eq_1}), we derive the following equation for $\hat u_x^\prime$
\begin{equation}
\label{eq_u_pr}
\frac{d}{d\tilde k_x} ( k^2 \hat u_x^\prime ) = - 2k_0 \left [ \frac{2-q}{q^2} (\arctan \tilde k_x - \arctan k_x) + \frac{k_x}{q k_0^2}  - \frac{\tilde k_x}{k^4} \right ].
\end{equation}
Given the zero initial condition $\hat u_x^\prime (t=0)=0$, we find that 
\begin{multline}
\label{u_pr}
k^2 \hat u_x^\prime = - \frac{2k_x}{q k_0} \left ( \tilde k_x - k_x \right ) +  
\frac{2k_0(2-q)}{q^2} \left [\tilde k_x\left ( \arctan k_x - \right . \right . \\ \left . \left . \arctan \tilde k_x \right ) + 
\ln \left (\frac{k}{|k_0|} \right )  \right ] 
+ \frac{1}{|k_0|} - \frac{|k_0|}{k^2}.
\end{multline}

We estimate $k_{z\,\max}$ from the assumption that it corresponds to a constant ratio of the amplitude of $k_z^2\hat u_x^\prime$ and the amplitude of leading order $\hat u_x$ 
given by equation (\ref{u_x_analyt})
\begin{equation}
\label{k_z_max_cond}
k_{z\, \max}^2 \frac{|k^2 \hat u_x^\prime|}{|k_x|} \Biggr\rvert_{\tilde k_x=\tilde k_{x\, {\ max}}} = \epsilon \ll 1,
\end{equation}
where $\tilde k_{x\, {\max}}$ is taken at some time after the instant of swing and corresponds to a certain location of SFH on its plateau.
Note that the higher is $|\tilde k_{x\, {\max}}|$, the smaller is $k_{z\, \max}$ and the longer is the duration of our analytical approximation. 

Most naturally, $\tilde k_{x\, {\max}}$ can be evaluated at the latest extremum of $\hat u_z(t)$ which emerges as one incorporates viscous damping into the problem.

\section{Taking viscous damping into account}
\label{sect_visc}

Viscous force, at first, confines the swing amplification of SFH as the initial winding of the spiral and so the duration of 2D transient growth become limited by viscous dissipation.
Thus, the largest value of $g_s$ corresponds to some $k_x = k_{x\,\max}$.
At second, viscous force causes damping of vertical motions and the asymptote (\ref{u_z_analyt_lim}) is replaced by extremum of $\hat u_z$ at some time after the instant of swing. Below we identify 
this time with $\tilde k_x = \tilde k_{x\,\max}$ used in equation (\ref{k_z_max_cond}), see Section \ref{sect_g_max}.

To determine both $k_{x,\max}$ and $\tilde k_{x\,\max}$ rigorously, let us suppose that $\hat u_x, \hat \omega_x$ 
is the solution of equations (\ref{rot_eq_1}-\ref{rot_eq_2}) obtained in the inviscid limit $R\to\infty$.
Then, the solution of the same equations taken with the account of non-zero viscosity $R<\infty$ is represented by
\begin{eqnarray}
\hat  u_{x,\,\nu} = \hat u_x {\rm e}^{-\gamma}, \nonumber \\
\hat \omega_{x,\,\nu} = \hat \omega_x {\rm e}^{-\gamma}, \nonumber
\end{eqnarray}
where 
\begin{equation}
\label{gamma}
\gamma = \frac{1}{qR} [ (1+k_z^2) (\tilde k_x - k_x) + \frac{1}{3} (\tilde k_x^3 - k_x^3) ].
\end{equation}
Note that for simplicity below we neglect by $k_z^2$ in comparison with unity in (\ref{gamma}) since $k_z\lesssim 1$ as far as $q \, \epsilon\lesssim 1$ which is assessed from the condition (\ref{k_z_max_cond}).

The spanwise velocity perturbation is modified in the same way
\begin{equation}
\label{u_z_visc}
\hat u_{z,\,\nu} = \hat u_z {\rm e}^{-\gamma},
\end{equation}
where $\hat u_z$ is given by equation (\ref{u_z_analyt}). This allows one to obtain $k_{x,\max}$ and $\tilde k_{x\,\max}$ jointly as the roots of the following set of equations
\begin{eqnarray}
\label{max_u_z_1}
\frac{\partial \hat u_{z,\,\nu}}{\partial k_x} = 0, \\
\label{max_u_z_2}
\frac{\partial \hat u_{z,\,\nu}}{\partial \tilde k_x} = 0,
\end{eqnarray}
provided that the initial SFH is assumed to be the leading spiral ($k_{x\,\max} < 0$ and $\tilde k_{x\,\max} > 0$) in quasi-Keplerian flow ($q>0$) and, vice versa, the trailing spiral ($k_{x\,\max}>0$ and $\tilde k_{x\,\max}<0$) in cyclonic flow ($q<0$).

\subsection{The case of tightly wound SFH}

In the limit of $|k_{x\,\max}|\gg 1$ and $|\tilde k_{x\,\max}|\gg 1$, which is valid for sufficiently high $R$, the approximate relations 
$$\arctan \{ k_{x\,\max}, \tilde k_{x\,\max} \} \approx {\rm sgn}(q)\pi/2 - \{ k_{x\,\max}^{-1}, \tilde k_{x\,\max}^{-1} \}$$ can be substituted into equation (\ref{u_z_analyt}).
Then, equation (\ref{max_u_z_1}) simply recovers an approximate value of $k_{x\,\max}$, which follows from \citet{razdoburdin-zhuravlev-2017} estimation of maximum duration of the swing amplification $t_{\max}$ (see their equation B2):

\begin{equation}
\label{k_x_max}
k_{x\, \max} \approx -q \, t_{\max}= - {\rm sgn}(q) (R |q|)^{1/3}.
\end{equation}
Thus, a very high transient growth of 2D SFH $\propto R^{2/3}$, and so the amplitude of the cross-rolls, can be reached taking into account that the Reynolds number 
may be as huge as $10^{10}$ in astrophysical boundary layers and discs. 

Further, equation (\ref{max_u_z_2}) yields the following estimation for $\tilde k_{x\,\max}$ 

\begin{equation}
\label{tilde_k_x_max}
\tilde k_{x\,\max} \approx {\rm sgn}(q) \left ( \frac{2|q|R}{(2-q)\pi} \right )^{1/4}
\end{equation}
provided that $R^{1/2}\gg 1$.

\section{Maximum growth factor of cross-rolls}
\label{sect_g_max}

The growth of the cross-rolls can be measured by energy density stored in vertical motion. 
In this case, since the initial condition (\ref{init_cond}) sets SFH with the unit energy and the planar velocity field, the growth factor of the cross-rolls is defined as
\begin{equation}
\label{g_z}
g_z \equiv \hat u_{z,\,\nu}^2.
\end{equation}
Then, the quantity
\begin{equation}
\label{g_z_max}
g_{z\,\max} \equiv \hat u_{z,\,\nu}^2( k_{x\,\max}, \tilde k_{x\,\max}, k_{z\,\max} )
\end{equation}
can be considered as an estimate of the largest possible amplification of cross-rolls generated by shearing spirals. 
Note that $g_{z\,\max}/\epsilon$ is the function of $R$ and $q$.
One may assume that violation of our analytical approximation discussed in Section \ref{validity} corresponds to the breakdown of swing amplification being a
unique mechanism generating the cross-rolls. Of course, an exact $k_{z\,\max}$ corresponding to the cross-rolls of the largest amplitude can be obtained solving 
the full set of equations (\ref{rot_eq_1}-\ref{rot_eq_2}) without an expansion over $k_z$. 
Nevertheless, it is known that $k_{z\,\max}$ should be at least $k_{z\,\max}\lesssim |q/(2-q)|$\footnote{cf. equation (\ref{k_z_max}) below.}, as the parameter $\beta\gtrsim 1$ introduced by \citet{balbus-2006}, 
see their equation (39), substantially weakens the swing amplification and, consequently, the cross-rolls amplitude.
In this study we intend to draw main conclusions assuming that equation (\ref{g_z_max}) gives a reasonable estimate of the largest cross-rolls generated by shearing spirals, whereas  
their exact identification from strict solution of (\ref{rot_eq_1}-\ref{rot_eq_2}), as well as determination of the corresponding $k_{z\,\max}$, is subject for the future studies with the use of optimisation technique.

Below we determine the largest growth factor of cross-rolls using accurate and approximate procedures.
\begin{itemize}

\item[i)]

An accurate determination of $g_{z\,\max}(q,R)$ is done by solving equations (\ref{max_u_z_1}-\ref{max_u_z_2}) numerically, which subsequently provides us with $k_{z\,\max}$ from the condition (\ref{k_z_max_cond}).
Then, $g_{z\,\max}$ is obtained from equation (\ref{u_z_visc}) with $k_x = k_{x\,\max}$, $\tilde k_x = \tilde k_{x\,\max}$ and $k_z = k_{z\,\max}$.

\item[ii)]

In the case of sufficiently high $R$ $g_{z\,\max}$ can be estimated analytically. 
Indeed, as the conditions $k_{x\,\max}\gg 1$ and $\tilde k_{x\,\max}\gg 1$ are fulfilled, equation (\ref{u_pr}) simplifies to
$$
k^2 \hat u_x^\prime = -{\rm sgn}(q)\frac{2(2-q)}{q^2} |k_x| \tilde k_x
$$
and yields the following estimation for $k_{z\,\max}$
\begin{equation}
\label{k_z_max}
k_{z\,\max} \approx \left ( \frac{|q|^7}{2^5 \pi^3 (2-q)^3 R} \right )^{1/8} \epsilon^{1/2}.
\end{equation}
Finally, in order to estimate the largest growth factor of cross-rolls in this case, one must substitute equations (\ref{k_x_max}), (\ref{tilde_k_x_max}) and (\ref{k_z_max}) into 
\begin{equation}
\label{visc_max_u_z}
\hat u_{z,\,\nu} \approx \pi \, {\rm sgn}(q) k_{z} |k_{x}| \frac{2-q}{q} {\rm e}^{-\frac{\tilde k_{x}^3 - k_{x}^3}{3qR}}.
\end{equation}

\end{itemize}

\vspace{0.2cm}

\begin{figure}
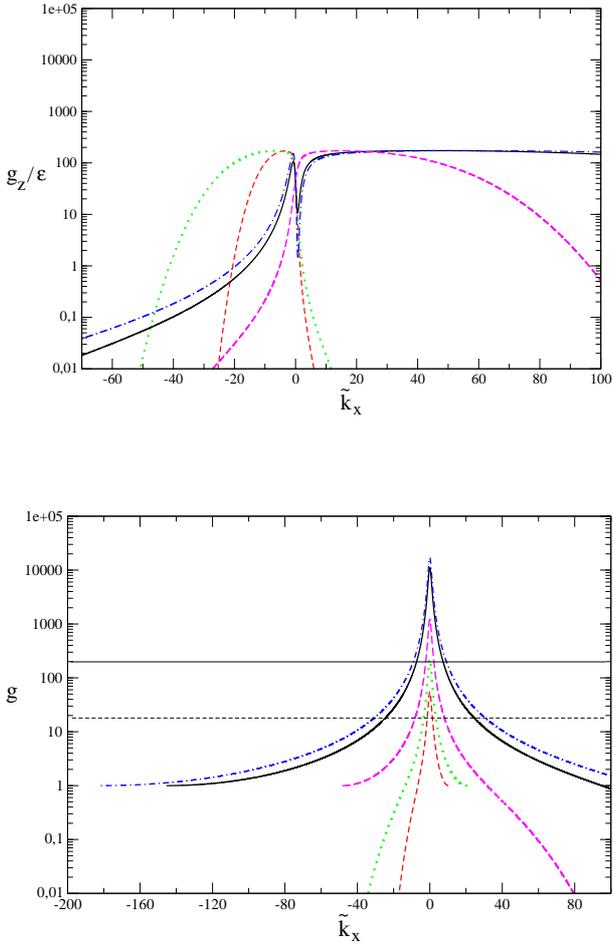

\begin{center}
\includegraphics[width=8cm,angle=0]{fig_1a}
\end{center}
\vspace{0.5cm}
\begin{center}
\includegraphics[width=8cm,angle=0]{fig_1b}
\end{center}
\caption{Short-dashed, dotted, long-dashed, solid and dot-dashed curves show $g_z(\tilde k_x)$ (top panel) and $g(\tilde k_x)$ (bottom panel) for transient growth of the particular SFH at $q=-5,-1.5, 0.5, 1.5$ and $1.6$, respectively. 
In each case, $k_x$ and $k_z$ are chosen in such a way that SFH attains the constant value $g_z(\tilde k_{x\,\max}) = g_{z\,\max}\approx 180\epsilon$, see Section \ref{validity} for definition of $\epsilon$.
This value of $g_{z\,\max}/\epsilon$ is obtained employing the procedure (ii) (see this Section) in order to fit the transitional dependence $R_T(q)$ found in the tall box simulations of turbulence in the cyclonic regime, 
see Section \ref{simulations} for details and the dashed curve on the top panel in Fig. \ref{fig:RT:Elongated} and in Fig. \ref{fig_8}. 
Solid and dashed horizontal lines on the bottom panel mark the value of corresponding $g_{z\,\max}$ for $\epsilon=1$ and $\epsilon=0.1$, respectively.
} \label{fig_1}
\end{figure}

\begin{figure}
\begin{center}
\includegraphics[width=8cm,angle=0]{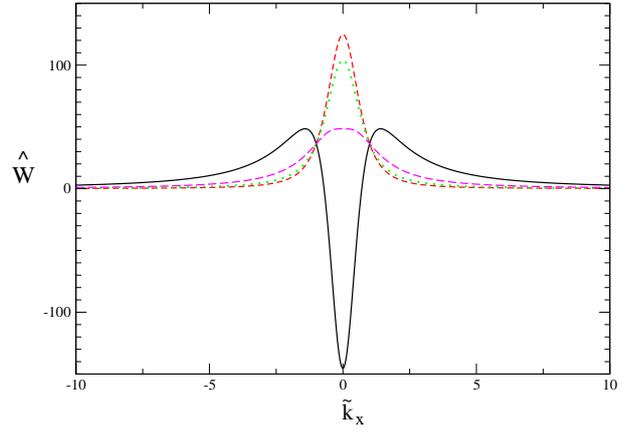}
\end{center}
\caption{Short-dashed, dotted, long-dashed and solid curves show $\Im [\hat W ](\tilde k_x)$ according to equation (\ref{W_analyt}) for $q=-5,-1.5, 0.5$ and $1.5$, respectively. 
The growth factors of the corresponding SFHs are shown in Fig.~\ref{fig_1}.
Note that there is no curve for $q=1.6$ since it virtually coincides with the curve for $q=1.5$.} \label{fig_2}
\end{figure}

Now, setting $g_{{z\,\max}}/\epsilon$ obtained according to the procedure (i) or (ii) to a constant value, we find $R(q)$ as a root of equation (\ref{g_z_max}).
This also implies that $k_{x\,\max}(q)$, $\tilde k_{x\,\max}(q)$ and $k_{z\,\max}(q)/\sqrt{\epsilon}$ have been determined,
so that we are ready to plot the growth factors of the particular SFH corresponding to constant $g_{z\,\max}/\epsilon$  for several values of $q$, see Fig. \ref{fig_1}.
These growth factors have to be regarded as functions of time or, equivalently, $\tilde k_x$: so $g(\tilde k_x)$ is defined by equation (\ref{TG_2D}) and $g_z(\tilde k_x)$ is defined by equation
(\ref{g_z}). As one can see on the top panel in Fig. \ref{fig_1}, all curves of $g_z(\tilde k_x)$ attain the same maximum value $g_{z\,\max}/\epsilon$, which
occurs after the instant of swing at negative and positive $\tilde k_x = \tilde k_{x\,\max}$ for cyclonic and quasi-Keplerian regimes, respectively. If it were not for viscous damping, $g_z$ would tend to the plateau, 
which has been found previously for the inviscid solution, see equation (\ref{u_z_analyt_lim}). 
While comparing two panels in Fig. \ref{fig_1}, it is seen that the ratio $(g_s/g_{z\,\max})\epsilon$, where $g_s$ is factor of 2D swing amplification of SFH, dramatically (and monotonically) increases as one goes from
large cyclonic shear rates across the solid-body line up to the marginal case of the Rayleigh line $q=2$. For example, at the Keplerian shear rate this ratio attains almost $10^2$.
Thus, the cross-rolls of the same intensity are generated by SFH with much larger $|k_x|$ as one approaches the Rayleigh line. The larger is $|k_x|$, the tighter is winding of the initial shearing spiral, the more 
it susceptible to viscous dissipation at a typical time $\propto |k_x|^{-2}$. Consequently, the closer one approaches the Rayleigh line, as starting from 
$q=-\infty$, the higher Reynolds number is required in order to generate cross-rolls of the same intensity. This is reflected in fit of our simulations of the subcritical transition to turbulence in the cyclonic regime, 
see Section \ref{simulations} and solid curve on the top panel in Fig. \ref{fig:RT:Elongated}, as well as in our prediction of the subcritical transition to turbulence in quasi-Keplerian regime, see Section \ref{prediction} 
and curve in Fig. \ref{fig_8}.

Next, Fig. \ref{fig_2} illustrates the immediate cause of suppression of the cross-rolls generated in the low-shear ($|q|\sim 1$) cyclonic regime 
and even more in the quasi-Keplerian regime in comparison with the high-shear ($|q|\gg 1$) cyclonic regime. 
There are the profiles of pressure perturbation amplitude as function of $\tilde k_x$ plotted for shearing spirals shown in Fig. \ref{fig_1}. 
For SFH these profiles also represent the behaviour of spanwise acceleration of fluid elements in the perturbed flow. At first, as the shear rate goes from 
$-\infty$ up to $1/2$, the net acceleration impulse in the spanwise direction decreases just because $|\hat W|$ in the vicinity of the instant of swing
attains a smaller value. Next, at $q=1/2$ the point $|\hat W|(\tilde k_x=0)$ transforms from maximum to minimum and, further, for $q>1$, 
$\hat W$ changes sign in the vicinity of the instant of swing. Eventually, as $q\to 2$, the net acceleration of fluid elements in positive and negative 
spanwise directions balance each other, so that one is left with planar velocity perturbation as the shearing spiral is wound up by the shear after the instant of swing.
Certainly, an additional dip in growth of the cross-rolls, which must occur in the vicinity of the solid-body line $q=0$, is imposed on the described picture.
Therefore, there must exist a local maximum of growth of the cross-rolls generated by the initial SFH with constant $k_x$ and $k_z$ in the quasi-Keplerian regime.
In Section \ref{prediction} we show that this local maximum comes out through the local minimum of the transition Reynolds number occurring near the $q=1/2$.

\subsection{Cross-rolls in comparison with rolls and streaks}
\label{sect_compar}

Let us estimate the growth factor of rolls generated by the anti-lift-up effect and streaks generated by the lift-up effect in centrifugally stable rotating flow. 
This is to be done in order to evaluate the role of those known mechanisms relative to the generation of cross-rolls discussed above. 

Both lift-up and anti-lift-up mechanisms describe the transient growth of axisymmetric perturbations. 
Consequently, we set $k_y=0$ in equations (\ref{orig_eq_1}-\ref{orig_eq_4}) and obtain the following set of equations for $\hat u_x$ and $\hat\omega_x$
\begin{equation}
\label{axisymm_u_x}
\frac{d \hat{u}_x}{dt}=-2\frac{k_z}{k_x^2+k_z^2}\hat{\omega}_x,
\end{equation}
\begin{equation}
\label{axisymm_om_x}
\frac{d \hat{\omega}_x}{dt}=(2-q) k_z u_x
\end{equation}
in the inviscid limit $R\to\infty$,
which yields the following equation for $\hat{u}_x$
\begin{equation}
\label{eq_u_x}
\frac{d^2 \hat{u}_x}{d^2 t}+ \sigma^2 \hat{u}_x = 0,
\end{equation}
where $$\sigma^2 \equiv \frac{2(2-q)}{1 + \alpha^2}$$ with $\alpha \equiv k_x/k_z$.

The solution of equation (\ref{eq_u_x}) reads
\begin{equation}
\label{u_x_solution}
\hat{u}_x = A \sin (\sigma t +\varphi ),
\end{equation}
where amplitude $A$ and phase $\varphi$ are to be determined from initial conditions.

Further, $\hat{\omega}_x$ is obtained with the help of equation (\ref{axisymm_om_x})
\begin{equation}
\hat{\omega}_x = -A(2-q)k_z \sigma^{-1}\cos (\sigma t+\varphi),
\end{equation}
whereas
\begin{equation}
\hat u_y = A (2-q) \sigma^{-1} \cos (\sigma t+\varphi ),
\end{equation}
since $\hat \omega_x = k_z \hat u_y$ for axisymmetric perturbations.
Finally, 
\begin{equation}
\hat{u}_z = -\alpha \hat{u}_x=-A \alpha \sin (\sigma t +\varphi ).
\end{equation}

\subsubsection{Growth of rolls}
\label{TG_rolls}

The lift-up mechanism provides the transient growth of streamwise rolls, i.e. initial vortices with $\hat{u}_y=0$.
This implies that $\hat{u}_x=\sqrt{1/(1+\alpha^2)}$ and $\hat{u}_z=-\sqrt{\alpha^2/(1+\alpha^2)}$ for perturbations with initial $\hat u_x^2 + \hat u_z^2=1$.
Thus, $A=\sqrt{1/(1+\alpha^2)}$ and $\varphi=\pi / 2$.

Consequently, the growth factor of rolls defined as $\bar g \equiv g + g_z$ reads
\begin{equation}
\bar g = \frac{2-q}{2}\sin^2 \sigma t + \cos^2 \sigma t.
\end{equation}
The corresponding maximum growth factor reads
\begin{equation}
\label{g_max_rolls}
\bar g_{\max} = \frac{2-q}{2}
\end{equation}
and becomes unlimited as $q\to -\infty$, i.e. in the non-rotating shear flow.

\subsubsection{Growth of streaks}
\label{TG_streakes}

The anti-lift-up mechanism provides the transient growth of streamwise streaks, i.e. initial vortices with
$\hat{u}_x=0$, $\hat{u}_y=1$, $\hat{u}_z=0$.
Thus, in this case $A=\sqrt{2/\left[(2-q)(\alpha^2+1)\right]}$, $\varphi=0$.

The growth factor of streamwise streaks is
\begin{equation}
\bar g = \frac{2}{2-q}\sin^2 \sigma t + \cos^2 \sigma t.
\end{equation}
The corresponding maximum growth factor reads 
\begin{equation}
\label{g_max_streakes}
\bar g_{\max} = \frac{2}{2-q}
\end{equation}
and becomes unlimited as $q\to 2$, i.e. on the Rayleigh line. 
\vspace{0.2cm}

Hence, maximum transient growth factor attained through either the lift-up or anti-lift-up mechanism is independent of SFH wavenumbers. Since there is no low limit on their absolute values in the boundless shear,
either the rolls or the streaks can be amplified up to $\bar g_{\max}$ given, respectively, by equations (\ref{g_max_rolls}) and (\ref{g_max_streakes}) at any $R$. At the same time, the cross-rolls cannot be amplified stronger 
than value estimated by definition (\ref{g_z_max}) with $\hat u_{z,\,\nu}$ substituted from equation (\ref{u_z_visc}). One can see, that additionally to constraint coming from damping of the swing amplification, which is
$g_{\max} \approx k_{x\,\max}^2$, the growth of the cross-rolls is reduced by factor $\sim ((2-q)/q) k_z^2$ comparing to $\bar g_{\max}$. 
We conclude that strictly in the absence of rotation or, alternatively, on the Rayleigh line cross-rolls are generally weaker than streaks or, respectively, rolls. 
However, when $-\infty < q < 2$, the transient growth of both rolls and streaks is highly suppressed. 
For example, the growth of rolls and streaks drops below $\bar g_{\max} \sim 20$, which is equal to $g_{z\,\max}$ represented in Fig. \ref{fig_1} at $\epsilon=0.1$, for $q\gtrsim -40$ and $q\lesssim 1.9$, respectively.
That is why in both the cyclonic and the quasi-Keplerian regime of homogeneous boundless shear flow cross-rolls are the only high amplitude 3D perturbations that can be generated via the transient growth mechanisms.

\section{Transition to turbulence: cyclonic regime}
\label{simulations}

In this Section we consider 3D non-linear dynamics of perturbations in the cyclonic regime of uniform and boundless rotating shear flow. 
Our goal is to recognize the subcritical transition to turbulence towards smaller $|q|$ compared to the previous studies and to obtain the transition Reynolds number, $R_T$.
This is done by performing hydrodynamical numerical simulations of the evolution of initial finite amplitude perturbations in the shearing box approximation.
Importantly, we finally check how the cross-rolls maximum growth factor estimated above behaves along the dependence $R_T(q)$.

\subsection{Model for non-linear dynamics of perturbations and numerical details}
Numerical simulations were carried out using open-source code ATHENA for solving continuity and Euler equations (see \cite{stone-2008} and \cite{stone-gardiner-2010})

\begin{equation}
\label{continuity}
\frac{\partial \rho}{\partial t}+\nabla \cdot (\rho \mathbf{v})=0,
\end{equation}
\begin{equation}
\label{euler}
\begin{aligned}
\frac{\partial (\rho \mathbf{v})}{\partial t} + \nabla \cdot (\rho \mathbf{v}\mathbf{v} & + P)  = 2 q \rho \Omega_0^2 {\bf x} - 2 \Omega_0\times \rho \mathbf{v} + \\
& \nabla \cdot \left(\rho \nu_0  \nabla \mathbf{v}\right) + \nabla \cdot  \left( \frac{1}{3}\rho \nu \nabla \cdot \mathbf{v}\right),
\end{aligned}
\end{equation}
supplemented by 
\begin{equation}
\label{eos}
P=c_s^2 \rho,
\end{equation}
where $x,y,z$ and $t$ are the Cartesian coordinates and time used in equations (\ref{sys1}-\ref{sys4}); ${\bf v}$, $P$ and $\rho$ are, respectively, velocity, pressure and density of the perturbed flow.
In equations (\ref{continuity}-\ref{eos}) kinematic viscosity $\nu$ and isothermal speed of sound $c_s$ are constants and the full non-linear dynamics of perturbations is considered in a box of size 
$L_x$, $L_y$ and $L_z$ along $x$, $y$ and $z$ directions, respectively. The usual periodic and shearing-periodic boundary conditions are imposed, respectively, along $y,z$ and $x$ directions 
at each side of the box, see e.g. \citet{hawley-gammie-balbus-1995}.
Note that everywhere below $L_x = L_y = 1$ but $L_z\ge L_x,L_y$.

In all our simulations
\begin{equation}
\label{small_box}
\max\{L_{x,y,z}\} \lesssim H,
\end{equation}
where $H \equiv c_s / \Omega_0$. According to hydrodynamical equilibrium in accretion disc boundary layer, $H$ is a typical scaleheight of the flow. 
The condition (\ref{small_box}) ensures that we work with vortical dynamics of perturbations as long as vortical initial perturbations are imposed.

It is suitable to parametrize the problem by two dimensionless numbers: Mach number
\begin{equation}
\label{Mach_def}
M=\frac{|q| \Omega_0 L_y}{c_s}
\end{equation}
and Reynolds number
\begin{equation}
\label{Re_nl_def}
R_{nl}=\frac{\Omega_0 L_y^2}{\nu}.
\end{equation}

Note that $R_{nl}$ is different from $R$ used above to describe the dynamics of the cross-rolls in the boundless flow. The relation between both 
Reynolds numbers can be derived along the following line.
Suppose that there is SFH with dimensional shearwise wavenumber $K_y$. 
In the box it takes values
\begin{equation}
\label{K_y}
K_y = \frac{2\pi}{L_y} n, 
\end{equation}
where $n$ is natural number. On the other hand 
$$
K_y = L^{-1},
$$
as far as in analytical linear model of transient growth $k_y=1$. 
Using the definitions (\ref{Re_def}) and (\ref{Re_nl_def}), we find
\begin{equation}
\label{Re_and_Re}
\frac{R_{nl}}{R} = 4\pi^2 n^2
\end{equation}
for SFH with shearwise wavelength $n$ times shorter than the shearwise box size.

In all simulations the density of the unperturbed flow is set to $\rho_0=1$ and the box rotation frequency $\Omega_0=0.001$.
Setting $q$, $M$ and $R_{nl}$ allows to specify unperturbed value of $c_s$ and value of $\nu$, as well as
pressure in the unperturbed flow, $P_0$, through the equation of state (\ref{eos}). 
Numerical resolution is chosen to keep a cubic form of a grid-cell, i.e. the number of nodes in $x$ and $y$ directions is equal to each other $N_x=N_y \equiv N$ and the number of nodes in spanwise direction is $N_z = l_z N$, 
where the box ratio $l_z\equiv L_z/L_y$.
We used $N=64$ once in order to compare our results with that by \citetalias{lesur-longaretti-2005} for the case of $q=-10$.
In all other simulations $N=128$.

\subsection{Indication of transition to turbulence}
In order to recognize turbulence, we employ a dimensionless angular momentum flux
\begin{equation}
\label{alpha}
\alpha=-\frac{<\rho v_x v_y^{\prime}>}{\rho_0 c_s^2}
\end{equation}
arising due to velocity perturbation to the unperturbed value $v_y=-q\Omega_0 x$. 
In equation (\ref{alpha}) $v_x$ and $v_y^\prime$ introduce perturbations of shearwise and streamwise velocity components, respectively, and the brackets denote spatial averaging.

Angular momentum flux arising due to microscopic viscosity in unperturbed flow is the following
\begin{equation}
\label{alpha_lam}
\alpha_{\nu}=\frac{\nu_0 \rho_0 r d\Omega /dr}{P_0}=-\frac{M^2}{q R}.
\end{equation}

We compare $\alpha(t)$ with $\alpha_{\nu}$ in the course of the non-linear evolution of perturbations and interpret the instant enhanced transport of angular momentum 
$\alpha(t) > \alpha_{\nu}$ as the existence of turbulence at time $t$.

\begin{figure}
\includegraphics[width=\linewidth]{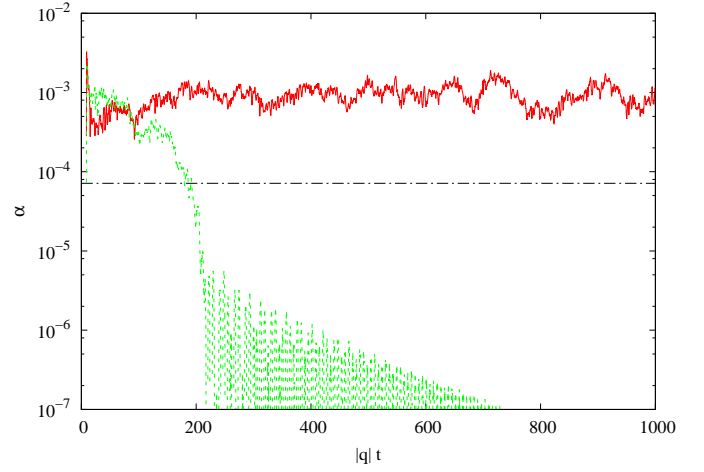}
\caption{
Behaviour of a dimensionless angular momentum flux for initial vortex of perturbations defined by eqs. (\ref{eq:ic4x}-\ref{eq:ic4z}) is shown for $q=-10$ and $M=1$.
Solid and dashed lines correspond to $R_{nl}=1400$ and $R_{nl}=1200$, respectively.
It was checked that turbulence for $R_{nl}=1400$ does not decay at least during the three thousands shear times $|q|t$.
Horizontal dot-dashed line corresponds to value of $\alpha_{\nu}$ in this case.}
\label{fig:alpha}
\end{figure}

Further, in order to check the transition to turbulence in the flow for given $R_{nl}$, the time-averaged variant of angular momentum flux is used also 
\begin{equation}
\label{alpha_t}
\tilde \alpha \equiv \frac{1}{t_2-t_1} \int \limits_{t_1}^{t_2} \alpha dt,
\end{equation}
where $|q| t_1=500\,\Omega_0^{-1}$ and $|q| t_2 = 1000\,\Omega_0^{-1}$. 
The transition to turbulence is defined by the condition
\begin{equation}
\label{indication}
\tilde \alpha > \alpha_\nu,
\end{equation}
which implies that self-sustained turbulent state exists for sufficiently long time.

\begin{figure}
\includegraphics[width=\linewidth]{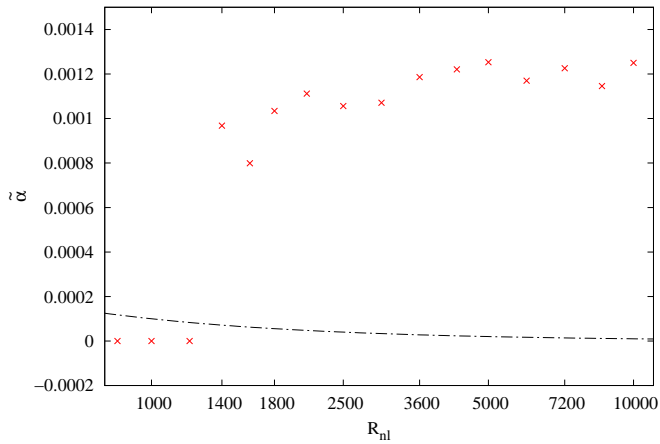}
\caption{
Time-averaged angular momentum flux is plotted as function of Reynolds number for $q=-10$ and $M=1$.
Dot-dashed line corresponds to value of $\alpha_\nu$ in this case.
The initial vortex of perturbations is determined by eqs. (\ref{eq:ic4x}-\ref{eq:ic4z}).
}
\label{fig:alphaAv}
\end{figure}

It was found that evolution of initial perturbations crucially depends on $R_{nl}$.
If $R_{nl}$ exceeds some threshold value, which we refer to as transition Reynolds number, $R_T$, initial perturbations give rise to self-sustained non-stationary solution.
In the opposite case $R_{nl}<R_T$, initial perturbations fade out, see Fig. \ref{fig:alpha}.
In order to determine $R_T$, we perform set of simulations on uniform grid of Reynolds number values.  
The transition Reynolds number is defined as $R_T=(R_{turb}+R_{damp})/2$, where $R_{turb}$ equals to the lowest $R_{nl}$ at which the flow demonstrates the transition to turbulence and 
$R_{damp}$ equals to the highest $R_{nl}$ at which we do not observe a long-living turbulent state of the flow and, consequently, the condition (\ref{indication}) is not satisfied.
As we see in Fig. \ref{fig:alphaAv}, the pass through $R_T$ is associated with a significant jump of $\tilde \alpha$.

\subsection{Initial conditions}

At first, we check whether the subcritical transition to turbulence we observe in our numerical experiment is in accordance with the results of \citetalias{lesur-longaretti-2005}.
Performing simulations in cubic box, they found $R_T = 1200$ at $q=-10$ for our definition of $R_{nl}$ 
(note that \citetalias{lesur-longaretti-2005} measure time in units of the inversed shear rate which makes their Reynolds number to be $R_{nl}$ multiplied by $q$).

Generally, this is a feature of the subcritical turbulisation that $R_T$ depends on shape and amplitude of the initial perturbations.
For example, this was shown experimentally by \cite{darbyshire-mullin-1995} and later numerically by \cite{faisst-eckhardt-2004} for pipe flow.
One may assume that for any particular values of $q$ and $M$ there are an optimal initial perturbations, which provide the lowest possible value of $R_T$.
These optimals can be found by variational approach to the non-linear Cauchy problem (see \cite{cherubini-2010b} and \cite{pringle-kerswell-2010} for such an example).
Unfortunately, this method is not easy to implement and it is time-consuming. In this work we restrict a comparison with \citetalias{lesur-longaretti-2005} by a single variant of initial perturbations. 
Also, we specify the same resolution as \citetalias{lesur-longaretti-2005}, $N_x=N_y=N_z = N \equiv 64$ using the cubic box.

We choose initial condition in the form of the following vortex
\begin{equation}
\label{eq:ic4x}
v_x=0,
\end{equation}
\begin{equation}
\label{eq:ic4y}
v_y^{\prime}=-\frac{A c_s \sqrt{2} K_z}{\sqrt{K_z^2+K_y^2}}\sin\left(K_y y+K_z z\right),
\end{equation}
\begin{equation}
\label{eq:ic4z}
v_z=\frac{A c_s \sqrt{2} K_y}{\sqrt{K_z^2+K_y^2}}\sin\left(K_y y+K_z z\right),
\end{equation}
where $K_y$ and $K_z$ are shearwise and spanwise dimensional wavenumbers of vortex and
the dimensionless constant $A$ specifies the initial specific kinetic energy of perturbations, which equals to $A^2/2$.

It was revealed that for $l_z=1$, $A=1$ and $K_y=K_z=2\pi$ the initial vortex (\ref{eq:ic4x}-\ref{eq:ic4z}) leads to self-sustained turbulence at $R_{turb} = 1400$ and damps at $R_{damp} = 1200$, see Fig. \ref{fig:alpha} 
and Fig. \ref{fig:alphaAv}. Thus, $R_T=1300$, which is close to value obtained by \citetalias{lesur-longaretti-2005}.
We additionally check that both the decrease and the increase of the amplitude of initial vortex, respectively, up to $A=0.5$ and $A=2$ leads to the increase of the transition Reynolds number, respectively, up to $R_T=2600$ and to $R_T=1500$.
Until the end of this Section we use the initial perturbations in the form  (\ref{eq:ic4x}-\ref{eq:ic4z}) with $A = 1$, $K_y = 2 \pi$ and $K_z = 2 \pi / l_z$.


\subsection{Results}

\begin{table}
\caption{
The summary of the numerical simulations for $M=1$, $N = 128$.
See text for definitions of variables.}
\label{tb:results}
\begin{tabular}{ccccc}
$q$     & $l_z$     & $R_{turb}$ & $R_{damp}$ & $\tilde \alpha\left(R_{nl}=R_{turb}\right)$\\ 
\hline
$-20$   & $1.0$     & $480$      & $400$      & $2.1 \times 10^{-3}$\\

\hline
$-17.5$ & $1.0$     & $630$      & $530$      & $1.8 \times 10^{-3}$\\

\hline
$-15$   & $1.0$     & $800$      & $670$      & $1.6 \times 10^{-3}$\\

        & $2.0$     & $500$      & $420$      & $1.6 \times 10^{-3}$\\

\hline
$-12.5$ & $1.0$     & $1000$     & $900$      & $1.2 \times 10^{-3}$\\

        & $2.0$     & $720$      & $600$      & $1.3 \times 10^{-3}$\\

\hline
$-10$   & $1.0$     & $1600$     & $1400$     & $7.7 \times 10^{-4}$\\ 

        & $2.0$     & $1200$     & $1000$     & $9.3 \times 10^{-4}$\\ 
\hline

$-9$    & $1.0$     &  $2000$    & $1700$     & $5.3 \times 10^{-4}$\\

        & $2.0$     &  $1400$    & $1200$     & $5.5 \times 10^{-4}$\\
\hline

$-8$    & $1.0$     &  $3000$    &  $2500$     & $4.0 \times 10^{-4}$\\

        & $2.0$     &  $2100$    &  $1800$     & $5.2 \times 10^{-4}$\\
\hline

$-7$    & $1.0$     &  $4900$    &  $4100$     & $2.2 \times 10^{-4}$\\

        & $2.0$     &  $3000$    &  $2500$     & $3.3 \times 10^{-4}$\\
\hline

$-6$    & $1.0$     &  $12500$   & $10500$     & $1.4 \times 10^{-4}$\\

        & $2.0$     &   $5000$   & $4300$      & $1.9 \times 10^{-4}$\\
\hline

$-5$    &  $1.0$    &  $43000$   & $36000$     & $4.0 \times 10^{-5}$\\

      	&  $2.0$    &  $10000$   & $8600$      & $1.2 \times 10^{-4}$\\
\hline

$-4$	& $1.0$     & $370000$   & $310000$    & $9.0 \times 10^{-6}$\\

     	& $1.125$   & $140000$   & $120000$    & $2.1 \times 10^{-5}$\\

     	& $1.25$    & $50000$    & $42000$     & $2.5 \times 10^{-5}$\\

     	& $1.5$     & $21000$    & $17500$     & $7.2 \times 10^{-5}$\\

     	& $2.0$     & $17500$    & $15000$     & $6.8 \times 10^{-5}$\\

     	& $3.0$     & $17500$    & $15000$     & $1.9 \times 10^{-5}$\\

     	& $4.0$     & $21000$    & $17500$     & $6.6 \times 10^{-5}$\\
\hline

$-3$	& $2.0$      & $36000$   & $43000$     & $1.4 \times 10^{-5}$\\

     	& $3.0$      & $36000$   & $43000$     & $8.5 \times 10^{-6}$\\
\hline
\end{tabular}
\end{table}

The picture of the subcritical transition to turbulence we extract from our simulations is presented in table  \ref{tb:results}.
For the first time (with regards to the model of unbounded uniform hydrodynamical flow considered here), we use not only the cubic box, but also a tall box with the box ratio $l_z>1$.
The box acts as a filter with respect to perturbations of different wavelengths. It prevents the existence of both small-scale perturbations with characteristic scale comparable and less than
the size of a numerical grid-cell and large-scale perturbations with characteristic scale larger than the size of the box in any direction. The tall box allows perturbation harmonics with
$K_z/K_y = l_z^{-1} < 1$ to take part in the non-linear evolution of the perturbed flow and, probably, in the subcritical transition to turbulence. 
In particular, this applies to shearing spirals which effectively generate the cross-rolls.
Indeed, employing equation (\ref{k_z_max}) one estimates 
$
K_z/K_y < k_{z\,\max}/\epsilon^{1/2} \approx 0.8
$
for $q=-10$ and $R_T=1300$ and becomes smaller as one shifts to smaller shear rate and higher Reynolds number. For example, at $q=-6.67$ and $R_T=10^4$ which is the least shear rate 
examined by \citetalias{lesur-longaretti-2005} for the subcritical transition, its value is less than approximately $0.5$. 
At the same time, the small size of the grid-cell is always required in order for the tightly wound spirals exhibiting large transient growth and generating high-amplitude cross-rolls to be involved into simulations.

The subcritical transition in cubic box has been observed up to $q=-4$, see table \ref{tb:results} and Fig. \ref{fig:RT:Elongated}.
This is a higher value comparing to $q=-6.67$ reached by \citetalias{lesur-longaretti-2005}, which is due to a higher numerical resolution used in our simulations.
Indeed, assuming that turbulisation, at least indirectly, is provided by the transient growth of shearing spirals and that the smallest shearwise scalelength of perturbations
resolved in the numerical scheme comprises 4 grid-cells, one can estimate the largest achievable $R_{nl}$  provided that the shearing spirals subject to the largest swing amplification
are yet resolved by the numerical scheme. The latter is done by equation (\ref{k_x_max})
\begin{equation}
\label{num_Re}
R_{nl} \approx 4\pi^2 \left ( \frac{N}{4} \right )^3 |q|^{-1}.
\end{equation}
For $N=128$ and $q=-4$ equation (\ref{num_Re}) yields $R_{nl} \approx 330 000$.
Note that this is close to $R_T$ obtained for $q=-4$, see table \ref{tb:results}, which naturally explains why we did not see the subcritical transition at smaller $|q|$.

\begin{figure}
\includegraphics[width=\linewidth]{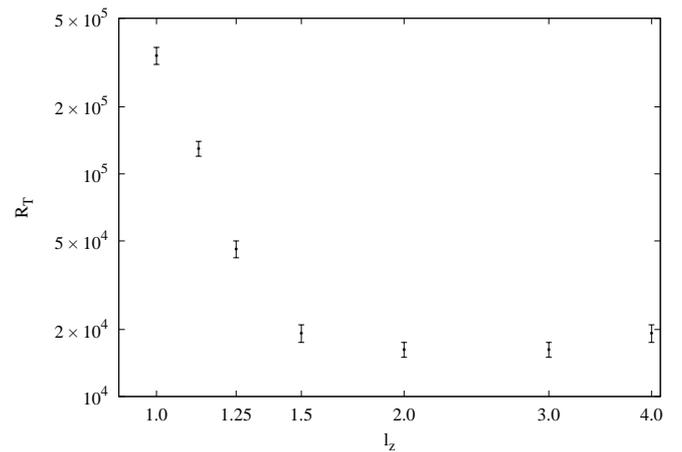}
\caption{
Transition Reynolds number $R_T$ vs. the box ratio $l_z$ is plotted for $M = 1$ and $q = -4$.
}
\label{fig:RT:Lz}
\end{figure}

\begin{figure}
\begin{center}
\includegraphics[width=\linewidth]{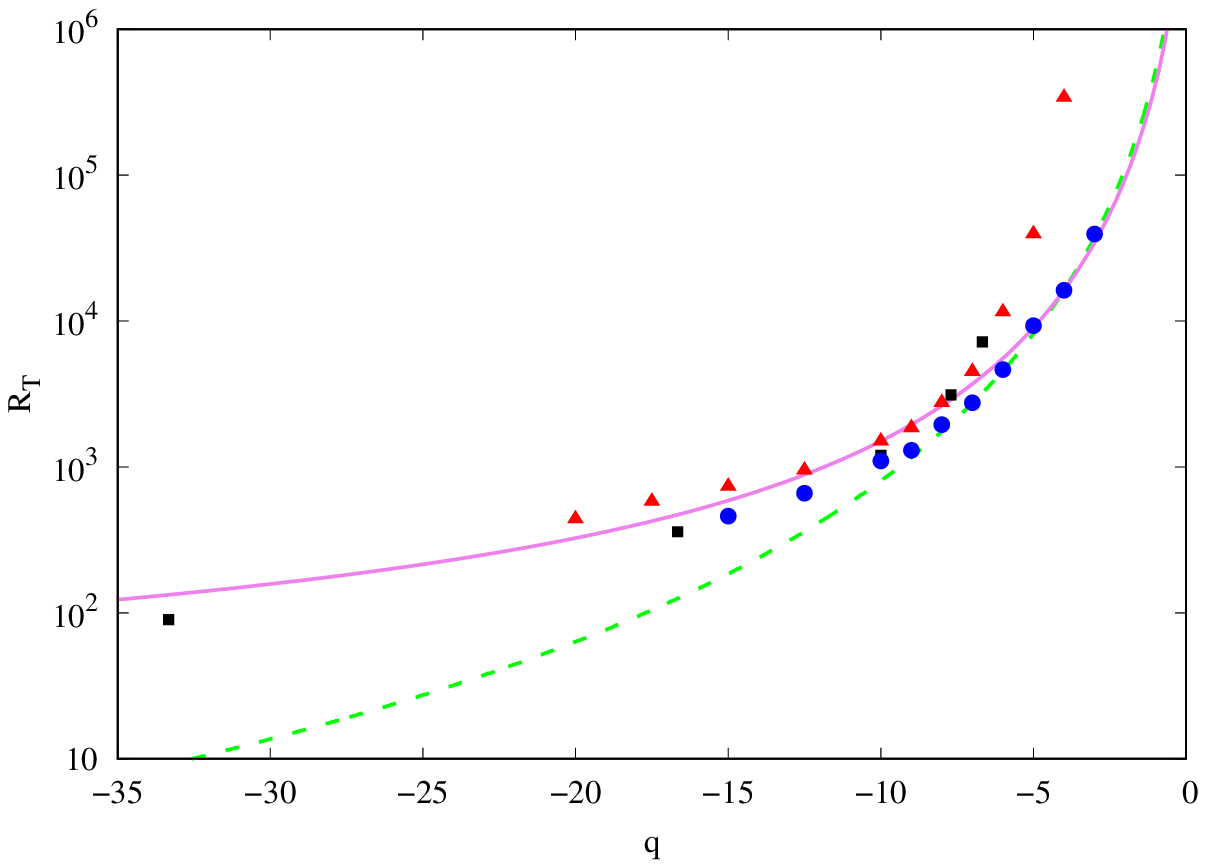}
\end{center}
\begin{center}
\includegraphics[width=\linewidth]{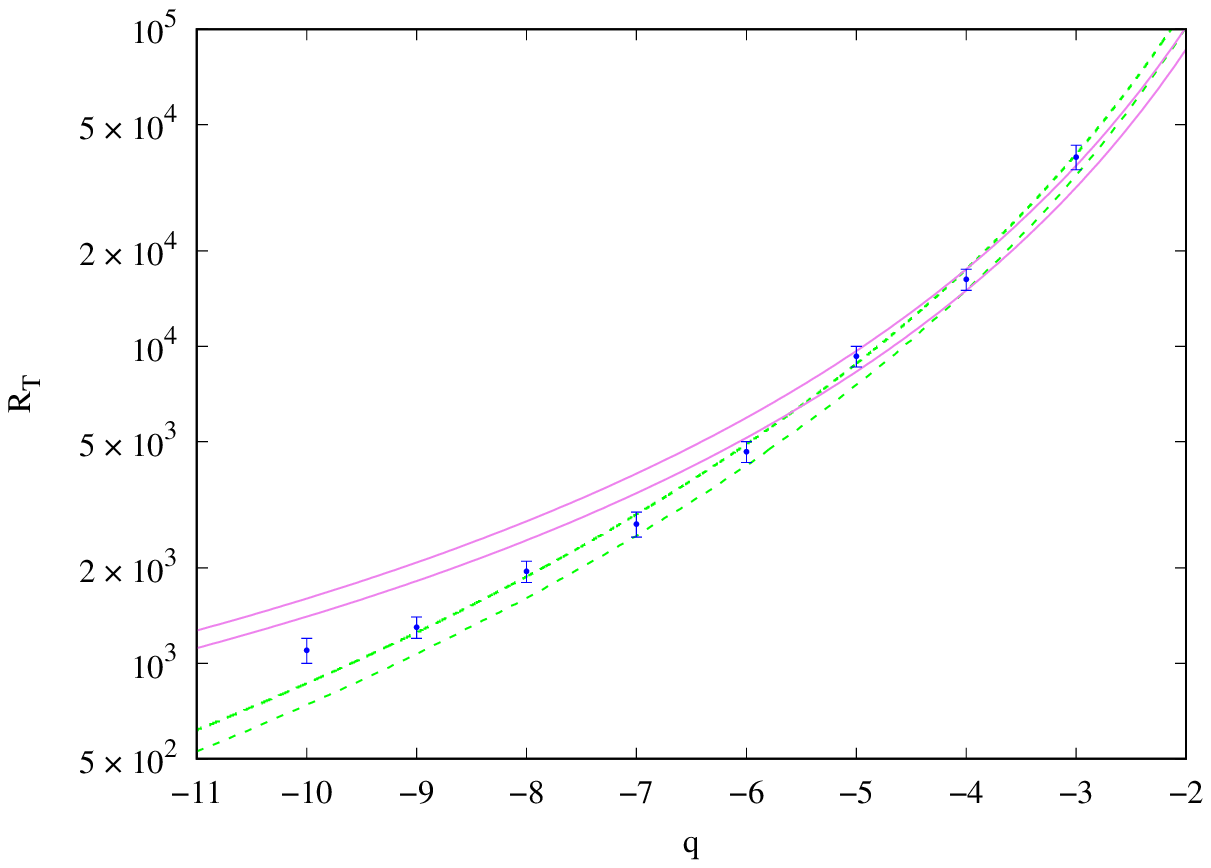}
\end{center}
\caption{
Transition Reynolds number $R_T$ vs. shear rate $q$ in the cyclonic regime.
Top panel:
(1) the squares represent DNS by \citetalias{lesur-longaretti-2005}, see their Fig. 4;
(2) the triangles represent DNS performed in this work employing cubic box;
(3) the circles represent DNS performed in this work employing tall box with box ratio $l_z=2$;
(4) the solid curve shows $R_{nl}(q)$ corresponding to constant maximum growth factor of the cross-rolls given by equation (\ref{g_z_max_eval}) and obtained using the procedure (i) introduced in Section \ref{sect_g_max};
(5) the dashed curve shows the same as described in (4), but with $g_{z\,\max}/\epsilon$ obtained using the procedure (ii) introduced in Section \ref{sect_g_max};
Bottom panel:
error bars for tall box DNS represented by circles on top panel are plotted in the range $q = -10 \div -3$ according to the values of 
$R_{damp}$ and $R_{turb}$ given in table \ref{tb:results}. 
The corresponding solid and dashed lines represent the same as on top panel but for $g_{z\,\max}/\epsilon$ evaluated for $R_T(q=-4)=15000$ (lower lines) and for $R_T(q=-4)=17500$ (upper lines).
}
\label{fig:RT:Elongated}
\end{figure}

However, once we fix $q=-4$ and perform set of simulations in the tall box, we find that $R_T$ falls down by factor $\sim 20$ as $l_z$ increases from $1.0$ up to $1.5\div 2$, 
see Fig. \ref{fig:RT:Lz}. We check that for $l_z > 2.0$ $R_T$ tends to horizontal asymptote. Further, we managed to find the subcritical transition at even higher $q=-3$
in the box with the same resolution $N=128$ but with the box ratio $l_z=2$. We regard this result as an evidence for crucial role of harmonics of perturbations with certain $0 < (K_z/K_y)_{T} < 1$\footnote{Since 
for such a small $|q|$ the obtained $R_T$ is close to the numerical restriction (\ref{num_Re}), this result is likely to be affected by the numerical viscosity. See 
appendix \ref{app_resolution} for the details.}.
It is plausible that $R_T(l_z)$ acquires an asymptotic value as soon as $l_z^{-1} \leq (K_z/K_y)_T$.
Note that such a harmonics already exists in the cubic box, however, it corresponds to $n>1$, see equation (\ref{K_y}), and, consequently, to the effective $R$ smaller by factor of $n^2$, see (\ref{Re_and_Re}). 
Harmonics of columnar shape corresponding to the highest effective $R$ ( i.e. to the largest shearwise wavelength equal to $L_y$), which implies the highest transient growth, emerges in tall box only.

The question arises, what can additionally testify in favor of cross-rolls as moderators of subcritical transition?
Trying to address this issue, we obtain $R_T$ in the tall box with $l_z=2.0$ for
various $q$, see table \ref{tb:results} and Fig. \ref{fig:RT:Elongated}. It was additionally checked that box ratio $l_z$ sufficient for $R_T(l_z)$ to approach the horizontal asymptote becomes smaller as one goes to higher $|q|$.
This guarantees that subcritical transition observed in the tall box with $l_z=2.0$ occurred in the presence of harmonics with $(K_z/K_y)_T$ in the range of the shear rate from $q=-15$ up to $q=-3$ covered by simulations.
It was revealed that the difference between $R_T$ obtained in cubic and tall boxes increases as $|q|$ becomes smaller.
Now, the tall box result $R_T = 16250$ at $q=-4$ allows us to evaluate the maximum growth factor of cross-rolls according to the procedure (i) formulated in Section \ref{sect_g_max} 
\begin{equation}
\label{g_z_max_eval}
g_{z\,\max}/\epsilon \approx 150.
\end{equation}
Note that previous inspection of the maximum optimal growth factor, $G_{\max}$, corresponding to transition to turbulence at $R_\Omega=0$ and $R_\Omega=-1$ 
in the framework of the shearing sheet model, see e.g. \citet{mukhopadhyay-2005}, has shown that $G_{\max}\sim 150$ in those cases, which coincides with 
(\ref{g_z_max_eval}) up to the smallness factor.
Using the procedure (i) further in order to look for the root of equation (\ref{g_z_max}), we obtain the curve of constant (\ref{g_z_max_eval}) on the plane $(R_T,\,q)$, see the solid line in Fig. \ref{fig:RT:Elongated}.
Additionally, the dashed line in Fig. \ref{fig:RT:Elongated} represents the corresponding approximate $R_{nl}(q)$ obtained according to the procedure (ii) formulated in Section \ref{sect_g_max}. 
A notable accordance is found between the growth of cross-rolls and the subcritical transition: namely, the largest growth factor of cross-rolls, as estimated in the main order in small $k_z$, 
remains almost constant along the transitional path, $R_T(q)$, deduced from our non-linear stability analysis. Below we will refer to (\ref{g_z_max_eval}) as the threshold value of $g_{z\,\max}/\epsilon$.
If there is a common non-linear mechanism sustaining the subcritical turbulence in uniform unbounded rotating shear flow of the cyclonic type, 
then it is natural to assume that it is triggered at the threshold value of the transient growth factor of relevant perturbations, at most, weakly 
changing with the shear rate. 
Here we find such an evidence in favour of the cross-rolls and their maximum growth factor, $g_{z\,\max}$, derived in the limit of small $k_z$.
It can be suggested that the subcritical transition, which occurs at asymptotic value of $R_T(l_z)$ in the tall box simulations, takes place due to a positive non-linear feedback from 
the harmonics of cross-rolls with $k_{z} \simeq (K_z/K_y)_T$. However, the ability of cross-rolls with $g_z \gtrsim g_{z\,\max}$ to generate secondary shearing spirals subject 
to swing amplification sufficient to sustain turbulence remains to be shown.

\begin{figure}
\begin{center}
\includegraphics[width=8cm,angle=0]{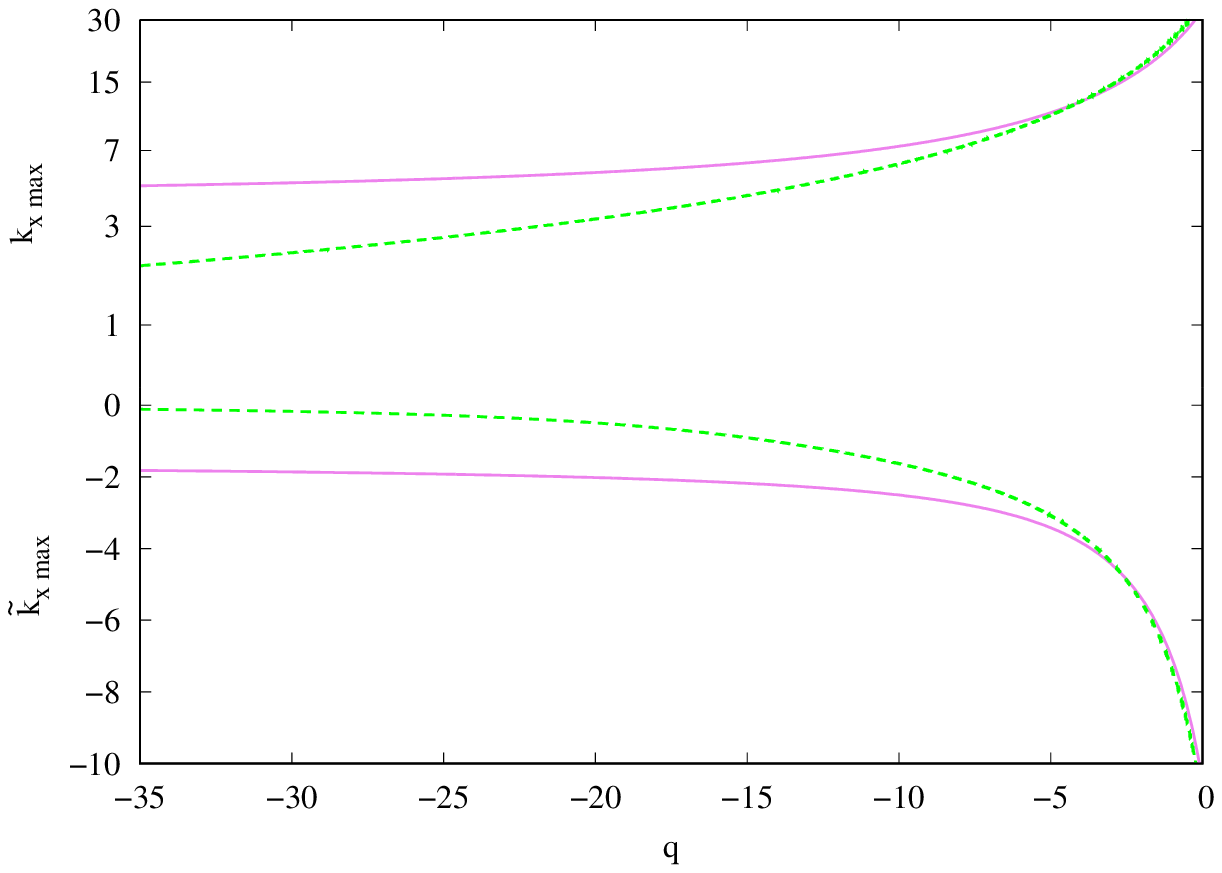}
\end{center}
\begin{center}
\includegraphics[width=8.2cm,angle=0]{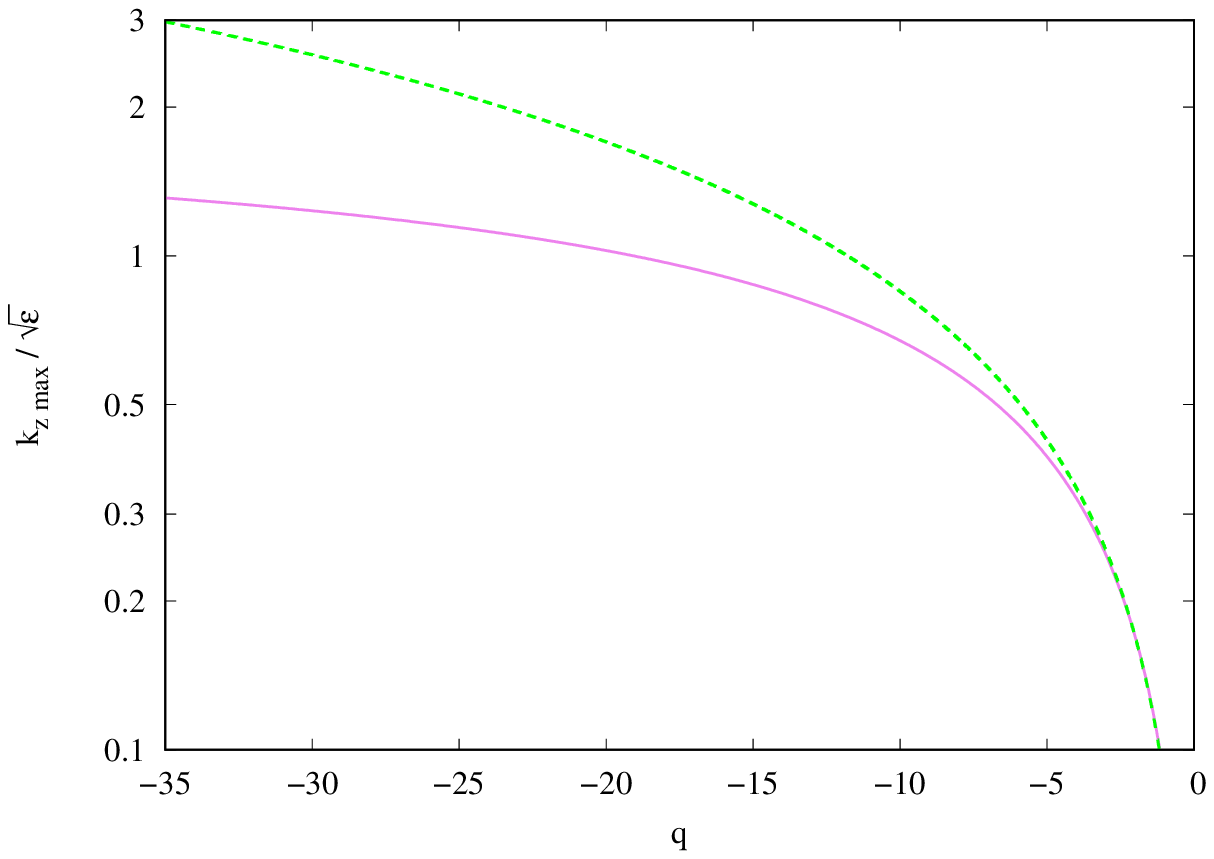}
\end{center}
\caption{Top panel: solid lines represent the roots of the set of equations (\ref{max_u_z_1}-\ref{max_u_z_2}) along $R_{nl}(q)$ plotted on the top panel in Fig. \ref{fig:RT:Elongated} by the solid line, while
dashed lines represent equations (\ref{k_x_max}) and (\ref{tilde_k_x_max}) along $R_{nl}(q)$ plotted on the top panel in Fig. \ref{fig:RT:Elongated} by the dashed line.
Bottom panel: solid line represents $k_{z\,\max}/\sqrt{\epsilon}$ obtained from the condition (\ref{k_z_max_cond}) with $k_x = k_{x\,\max}$ and $\tilde k_{x\,\max}$ along the corresponding solid lines on the top panel of this Figure,
while dashed line represents equation (\ref{k_z_max}) along $R_{nl}(q)$ plotted on the top panel in Fig. \ref{fig:RT:Elongated} by the dashed line.
}\label{fig_7}
\end{figure}

Note that as $q \lesssim -5$ the approximate analytical curve in Fig. \ref{fig:RT:Elongated} is no more in accordance with the curve obtained accurately. The reason for that is shown in Fig. \ref{fig_7}, where
$k_{x\,\max}$, $\tilde k_{x\,\max}$ and $k_{z\,\max}/\sqrt{\epsilon}$ along the corresponding curves of $R_{nl}(q)$ are plotted. Clearly, the limit of tightly wound spiral, $|k_{x\,\max}|, |\tilde k_{x\,\max}| \gg 1$,
is valid as far as $|q|\sim 1$ and breaks when $q \lesssim -5$. For $|q| \gg 1$ $R_{nl}(q)$ corresponding to constant (\ref{g_z_max_eval}) becomes too small for spirals to be tightly wound, see
equations (\ref{k_x_max}) and (\ref{tilde_k_x_max}).
At the same time, we find that the first point of the subcritical transition obtained by \citetalias{lesur-longaretti-2005} in the presence of rotation, $R_\Omega > 0$, which corresponds to $q\approx -33$, 
is also in reasonable accordance with the threshold value (\ref{g_z_max_eval}). Along with the comparison of rolls and cross-rolls growth factors with each other at $|q| \gg 1$, see the end of Section \ref{sect_compar}, 
this argues in favour of the assumption that, at least, as $q \gtrsim -35$, SSP is already replaced by a different self-sustaining solution incorporating the growth of cross-rolls. 

The dependence $k_{z\,\max}(q)$ obtained along the cross-rolls threshold growth factor (\ref{g_z_max_eval}), see Fig. \ref{fig_7}, confirms that the latter is exhibited by shearing spirals with large spanwise lengthscale.
As long as $|q|$ becomes smaller, the destructive role of Coriolis force in the evolution of SFH with non-zero $k_z$ increases as compared to the pressure gradients enabling their swing amplification. 
That is why $k_{z\,\max}$ decreases as one approaches the line of rigid rotation, which is seen in Fig. \ref{fig_7}, see also equation (\ref{k_z_max}). In turn, 
this naturally explains why the difference between $R_T$ obtained in cubic and tall boxes 
increases for smaller $|q|$: SFH with $n=1$, see equation (\ref{K_y}), corresponding to the growth factor (\ref{g_z_max_eval}) is
progressively suppressed by periodic boundary conditions imposed in the cubic box along spanwise direction.
Note that SFH that generate the cross-rolls of the largest amplitude will be highly columnar at $|q|\approx 1$, as far as $k_{z\,\max}/\sqrt{\epsilon}$ is estimated to become less than $0.1$. If subcritical turbulence in the cyclonic regime is sustained by the non-linear feedback from the cross-rolls,
one is worth checking its existence at $|q|\sim 1$ via the DNS in a very tall box.

\section{Transition to turbulence: towards the Keplerian shear across the solid-body line}
\label{prediction}

\begin{figure}
\begin{center}
\includegraphics[width=9cm,angle=0]{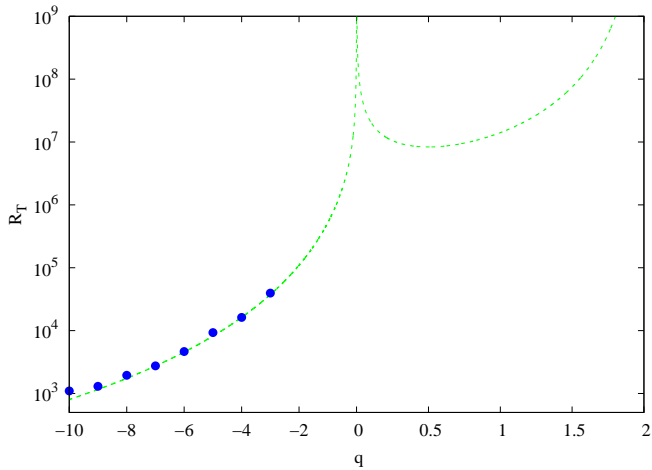}
\end{center}
\caption{The curve shows $R_{nl}(q)$ along the constant maximum growth factor of the cross-rolls given by equation (\ref{g_z_max_eval}), 
where $g_{z\,\max}/\epsilon$ is obtained approximately using the procedure (ii) introduced in Section \ref{sect_g_max}.
The circles represent DNS performed in this work using tall box with box ratio $l_z=2$, see Section \ref{simulations} for details.
Note that the curve is matched to the circle at $q=-4$, which corresponds to $R_T = 16250$, see table \ref{tb:results}.
} \label{fig_8}
\end{figure}

Let us suppose that cross-rolls are responsible for the subcritical transition in the whole range of centrifugally stable rotating shear flows.
If so, one can assume that corresponding threshold value of the cross-rolls largest growth factor, $g_{z\,\max}$, found in the cyclonic regime (see the previous Section) 
remains independent of the shear rate further in the quasi-Keplerian regime. 
Thus, we suppose that equality (\ref{g_z_max_eval}) holds on for $R_{nl} = R_T$ across the solid body line $q=0$ up to the Rayleigh line $q=2$.
Since the shearing spirals, which produce the threshold growth factor, are getting more and more wound up as one goes from $q=-\infty$ to $q=2$, see the bottom panel in Fig. \ref{fig_1} and the top panel in Fig. \ref{fig_7}, 
it is sufficient to use an approximate evaluation of $g_{z\,\max}$ provided by the procedure (ii) described in Section \ref{sect_g_max}.
So $R_{nl}(q)$ corresponding to a constant threshold growth factor (\ref{g_z_max_eval}) extends to quasi-Keplerian regime as shown by the curve in Fig. \ref{fig_8}.
This curve may be considered as the prediction of the subcritical transition to turbulence in homogeneous boundless quasi-Keplerian flow.  
Generally, the turbulence provided by the non-linear feedback from the cross-rolls is expected to occur at much higher Reynolds number $\sim 10^7\div 10^9$ in quasi-Keplerian regime when compared to cyclonic regime.
On the one hand, this is consistent with negative results from all existing studies of the non-linear instability of quasi-Keplerian regime, whereas on the other, 
even such a huge Reynolds number is feasible in astrophysical discs, see e.g. \citet{balbus-2003}.
Equations (\ref{k_z_max}) and (\ref{u_z_analyt_lim}) show that such an asymmetry of growth of the cross-rolls in cyclonic and quasi-Keplerian regimes is controlled by the factor
$$
g_{z\,\max} \propto (2-q)^{5/4},
$$
which stems from the suppression of fluid vertical motion in the course of the swing amplification of SFH, see comments to Fig. \ref{fig_2}. Consequently, we predict an asymmetry of $R_T(q)$ with respect to 
the line of rigid rotation: for the same absolute value of the shear rate, $|q|$, the subcritical transition in the quasi-Keplerian regime should occur at the substantially higher $R_{T}$ rather than in the cyclonic regime.

Let us note that the enhanced non-linear stability of quasi-Keplerian flow follows from our DNS in the cyclonic regime together with
the results of \citetalias{lesur-longaretti-2005} and \citet{rincon-2007} discussed in Introduction.
Indeed, the curves on top panel in Fig. \ref{fig:RT:Elongated} indicate that $R_T\approx 3\times 10^5$ at $q=-1.5$.
At the same time, from extrapolation of DNS by \citetalias{lesur-longaretti-2005} carried out at $R_\Omega < -1$, see expressions in their Fig.~7, it is expected that $R_T \gtrsim 10^9$ at $q=1.5$.
This independently shows an asymmetry of $R_T(q)$ with respect to change of sign of $q$.
However, the swing amplification itself produces maximum growth factor $\sim k_{x\,\max}^2$, which would yield a symmetric $R_T\propto |q|^{-1}$ 
with respect to change from cyclonic to quasi-Keplerian shear rate.

It is noteworthy that prediction of $R_T$ in Fig. \ref{fig_8} combined with extrapolations of DNS suggested by \citetalias{lesur-longaretti-2005} for quasi-Keplerian regime gives the most stable rotating flow located 
at super-Keplerian shear rate $q\sim 1.7 \div 1.8$.

Further, since in the vicinity of the solid-body line $R_T$ naturally tends to infinity (cf. equations (\ref{visc_max_u_z}) and (\ref{k_x_max}), (\ref{k_z_max}) at $|q|\to 0$), 
a minimal $R_T$ is predicted for quasi-Keplerian regime. The corresponding, most 'favourable' value of $q$ can be immediately estimated in the limit of tightly wound SFH. 
Using equation (\ref{visc_max_u_z}) in combination with equations (\ref{k_x_max}) and (\ref{k_z_max}) and the condition $R\gg 1$, we find that constant $g_{z\,\max}$ corresponds to
$$
(R_T q)^{2/3}(2-q)^2 = {\rm const},
$$
which yields minimal $R_T$ at $q=1/2$. So, $R_T(q=1/2)$ exceeds $10^7$, while the transition at the Keplerian shear, $q=3/2$, 
may occur at $R$ approaching $\sim 10^8$. Hence, the super-Keplerian flow $3/2<q<2$ is predicted to be the most non-linearly stable throughout the 
whole range of centrifugally stable rotating shear flows. Notably, most of the efforts to detect turbulence in quasi-Keplerian regime in laboratory experiments and DNS 
has been made at super-Keplerian shear rates. In contrast, our results suggest that one should extend the turbulent solutions obtained so far in the cyclonic
regime to smaller $|q|$ and then across the solid-body line, finally focusing the efforts on sub-Keplerian rotation.
The effective numerical viscosity existing in DNS is approximately proportional to the distance between the nodes of the numerical grid. 
If so, the maximum Reynolds number accessible in DNS is $ \propto N^2$.
Then, assuming that $R_T \approx 40000$ at $q = -3$, see table \ref{tb:results}, is the maximum accessible $R_{nl}$ for $N=128$ in tall box,
we suppose that in order to be able to check hypothetical transition to hydrodynamic turbulence in quasi-Keplerian flow,
at least at 'favourable' $q=1/2$, $N\gtrsim 2000$ is required. This is quite a high DNS resolution, however, it is not unreachable for the most powerful modern supercomputers.

\section{Summary}

\begin{figure*}
\begin{center}
\includegraphics[width=0.3\textwidth,angle=-90]{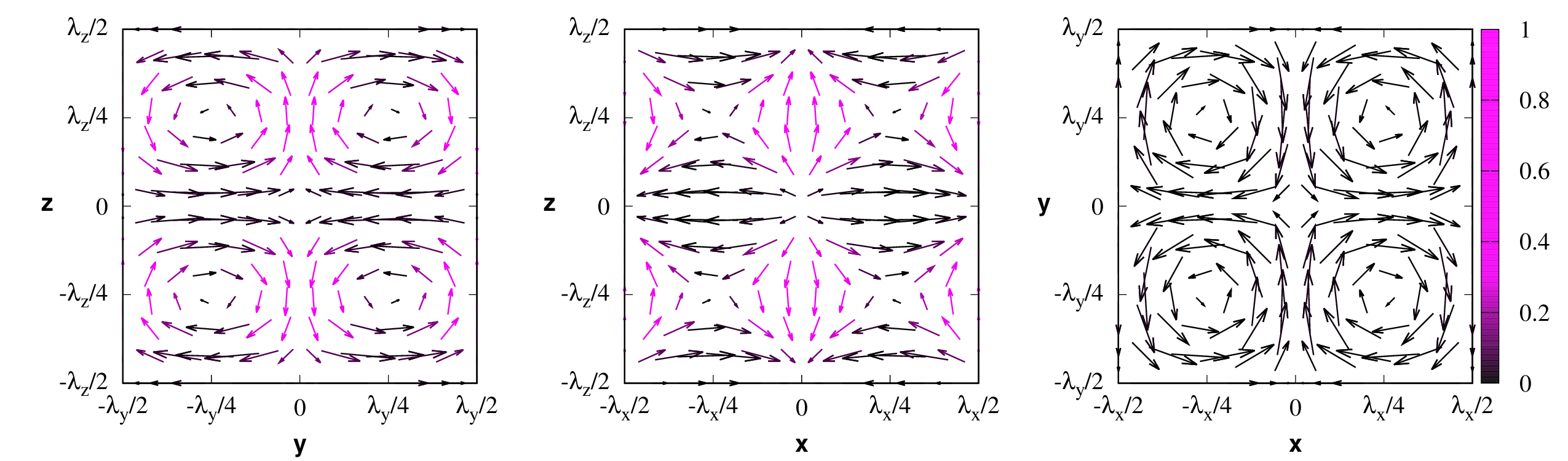}
\end{center}
\begin{center}
\includegraphics[width=0.3\textwidth,angle=-90]{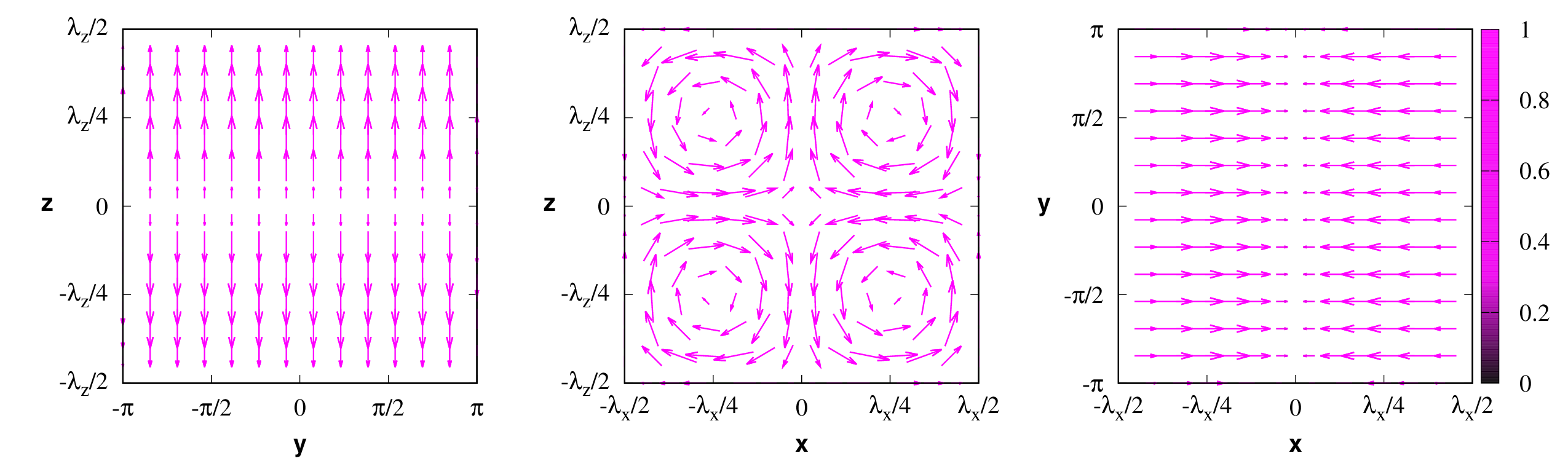}
\end{center}
\caption
{
Left, middle and right panels represent the cross-sections of shearing spiral velocity pattern, respectively, in $(y,z)$-, $(x,z)$- and $(x,y)$-planes, which 
contain the point $x=y=z=0$. The colour represents the velocity absolute value normalised by its maximum value 
separately for top and for bottom panels.
On top panels:
$k_x=-141$, $k_z=0.1$, $\tilde k_x = 43$ and $q=1.5$.
The wavelengths are $\lambda_x=2 \pi / \tilde k_x$, $\lambda_y=2 \pi$, $\lambda_z=2 \pi / k_z$.
On bottom panels:
$k_x=-141$, $k_z=0.1$ and $t=\pi / (2 \sigma)$, where $\sigma$ is defined in Section \ref{sect_compar};
the shear rate is $q=1.99$.
The wavelengths are $\lambda_x=2 \pi / k_x$, $\lambda_y=\infty$, $\lambda_z=2 \pi / k_z$. Extraction of shearing spiral velocity pattern is described in detail in Appendix \ref{app_represent}.
}
 \label{fig_4_5}
\end{figure*}

Angular momentum transport in boundary layers around the accreting weakly magnetised stars may be produced by turbulent fluctuations. 
Here we suggest that subcritical turbulence in boundless and homogeneous rotating shear flow of the cyclonic type, which is local representative of the boundary layer, is supplied with energy by nearly optimal shearing spirals
which generate essentially 3D vortices of high amplitude as a 'byproduct' of swing amplification. We call those 3D vortices 'cross-rolls' in order to distinguish them from the known streamwise rolls generated via anti-lift-up mechanism in a flow
on the Rayleigh line. Such turbulence must be sustained by a novel type of positive non-linear feedback from the cross-rolls, which would recover basin of growing shearing spirals. 
Furthermore, we suppose that this scenario also works in quasi-Keplerian regime.
These suggestions are supported by the following set of evidence.

\begin{itemize}
 \item[i)]
 
As well as known spanwise invariant (columnar) shearing spirals, the shearing spirals with small but non-zero spanwise wavenumber are subject to swing amplification, 
which causes their transient growth almost up to the highest possible values corresponding to growth factor much greater that unity. 
Thus, these vortices satisfy a necessary condition for the subcritical transition in both cyclonic and quasi-Keplerian regimes.
At the same time, it is known that all shearing harmonics with $k_z\gtrsim 1$ do not exhibit considerable transient growth since they drastically degenerate into epicyclic oscillations.

\vspace{0.1cm}

\item[ii)]

The substantial decrease of $R_T$ and the existence of an asymptotic value of $R_T(l_z)$ for large $l_z$ found in the tall box simulations, see Fig. \ref{fig:RT:Lz}, 
indicates that these perturbations are crucial for the transition and sustenance of subcritical turbulence, at least, in the cyclonic regime.
This is the most striking evidence that the cross-rolls should be ingredients of positive non-linear feedback completing the bypass scenario in
centrifugally stable rotating shear flows.

\vspace{0.1cm}

\item[iii)]

It is revealed that the maximum growth factor of cross-rolls decreases as the shear rate alters from $q=-\infty$ through the solid-body line $q=0$ up to $q=2$, see Fig. \ref{fig_1}.
Moreover, its constant value accords with the curve of $R_T(q<0)$ obtained in tall box simulations, see Fig. \ref{fig:RT:Elongated}. 
We regard this value as the threshold growth factor, which triggers the non-linear sustenance of subcritical turbulence in the cyclonic regime.

On the contrary, the spanwise invariant shearing spirals with $k_z=0$ do not produce any 3D vortices and acquire growth factor
symmetric with respect to change between cyclonic and quasi-Keplerian regimes at the same $|q|$. 
However, DNS of turbulence in the cyclonic regime performed by \citetalias{lesur-longaretti-2005} 
and in this work far beyond the zone of the lift-up effect show strong variation of 2D growth factor along the obtained $R_T(q<0)$.

\vspace{0.1cm}

\item[iv)]

The disappearance of the cross-rolls in the vicinity of $q=2$ naturally explains the asymmetry of the transition to turbulence revealed in DNS close to 
$R_\Omega=-1$ versus $R_\Omega=0$: the bypass scenario based on the transient growth of shearing spirals with small but non-zero $k_z$ for $q<0$ replaces SSP, 
retaining the non-linear instability of the cyclonic flow immediately after the lift-up ceases to operate, 
whereas a similar situation with the anti-lift-up mechanism is not realised in the case of quasi-Keplerian rotation.

\vspace{0.1cm}

\item[v)]

The cross-rolls are the only high-amplitude essentially 3D vortices that can be generated via the transient growth mechanism 
in the whole range of centrifugally stable shear rates provided that the Reynolds number is sufficiently high. 
They create the vorticity (namely, its shearwise and streamwise components), which must
be of crucial importance for 3D turbulence.

\vspace{0.1cm}

\item[vi)]

The transition to turbulence in the tall box is observed until smaller $|q|$ than in the cubic box at the same numerical resolution, see. Fig. \ref{fig:RT:Elongated}. Generally, a sensitivity of the transition to shape of a box indicates about anisotropic nature of subcritical turbulence.

In the future work its properties including the details of the non-linear sustaining process, as well as the precise role of the cross-rolls in it, 
should be studied employing, in particular, the 
3D Fourier analysis in ${\bf k}$-space, see \citet{mamatsashvili-2016} and \citet{mamatsashvili-2017} for an examples of such studies in application
to non-rotating shear flow and Keplerian flow with an azimuthal magnetic field.

\end{itemize}

The shearing spirals considered analytically in the leading order in small $k_z$, allow to draw the following conclusions.

\begin{itemize}

\item[i)]

Dynamics of perturbations in the plane orthogonal to spanwise direction remains approximately the same as for spanwise invariant (columnar) shearing spirals, 
which is the 2D swing amplification process.

\vspace{0.1cm}

\item[ii)]

Spanwise acceleration of fluid elements is determined by the swing amplification.

\vspace{0.1cm}

\item[iii)]

Spanwise velocity perturbation increases while the shearing spiral swings from trailing to leading in the cyclonic flow 
and vice versa in the quasi-Keplerian flow. After that, the spiral is wound by the shear, which causes the decay of streamwise and shearwise velocity perturbations, 
whereas the spanwise velocity perturbation tends to asymptotic inviscid non-zero value proportional to the net impulse given to fluid elements in spanwise direction. 
The latter depends on the shear rate and vanishes as $q\to 2$. Viscous damping results in a maximum value of the initial winding of shearing spiral 
and additionally in the maximum of spanwise velocity, which appears after the instant of swing, see equations (\ref{k_x_max}) and (\ref{tilde_k_x_max})
giving the locations of those extrema in the limit of tightly wound spirals.

\vspace{0.1cm}

\item[iv)]

Assuming that the cross-rolls of the largest amplitude sustain turbulence throughout the range of centrifugally stable rotating shear flows 
and the threshold value of their growth factor found in the cyclonic regime, see equation (\ref{g_z_max_eval}), remains the same,   
we obtain a tentative prediction of $R_T(q)$ for the quasi-Keplerian regime, where hydrodynamical turbulence is not yet discovered. 
In particular, we find that the lowest $R_T\sim 10^7$ in the quasi-Keplerian regime must be attained at sub-Keplerian shear rate $q\approx 1/2$. 

\item[v)]

The threshold value of the cross-rolls growth factor (\ref{g_z_max_eval}) exceeds the maximum growth factors of streaks and rolls generated via lift-up
and anti-lift-up up to the shear rates very close, respectively, to non-rotating flow and to flow on the Rayleigh line, see Section \ref{sect_compar}.

\end{itemize}

The cross-rolls appearance is shown on top panels in Fig. \ref{fig_4_5}, see also Appendix \ref{app_represent} for the explanation of this representation.
Note that the growth factor of corresponding shearing spiral is plotted by solid curve in Fig. \ref{fig_1}. To compare with, the rolls generated from axisymmetric streaks
via anti-lift-up with the same $k_x$ and $k_z$ but in the flow close to the Rayleigh line are shown on the bottom panels in Fig. \ref{fig_4_5}. 
The the former and the latter ones are aligned in the shearwise and the streamwise directions, respectively. Furthermore, in contrast to rolls, the cross-rolls are essentially 3D vortices, 
since they are of a finite length along their axes of rotation being contracted by the background shear $\propto |\tilde k_x|^{-1}$. 
In combination with the opposite rotation of the adjacent cross-rolls this causes monotonic growth of $\omega_y\propto |\tilde k_x|$ in areas between the
adjacent cross-rolls, in the vicinity of, e.g., the point $(x=\lambda_x/4,y=0,z=\lambda_z/4)$. As one can see, the cross-rolls coexist with 'planar' eddies
fading after the instant of swing. 
Such a complicated structure of perturbation velocity field requires a new concept of the non-linear feedback, which may be completely different from both SSP and 
a scheme associated with streamwise rolls in the flow on the Rayleigh line. Note that the latter itself remains unknown.

The prediction of $R_T(q)$ obtained in this work, see Fig. \ref{fig_8}, makes it clear that, at least for boundless homogeneous flows, 
the issue of the transition to turbulence in Keplerian regime should be attacked from the side of the cyclonic regime. This implies 
the crossing of solid-body line, $R_\Omega=+\infty \to R_\Omega=-\infty$, which is not evident while looking at axis of $R_\Omega$.
However, we suppose that $R_T \sim 10^7$ at the most 'favourable' sub-Keplerian shear rate $q \sim 1/2$, which presumably requires resolution
$N\gtrsim 2000$ for the spectral Fourier code to be able to test the non-linear instability of quasi-Keplerian regime. 
Still, such an extraordinary non-linear stability of Keplerian flow may be defeated by huge spatial scales of real astrophysical discs leading to 
$R= 10^{10} \div 10^{13}$. 


\section*{Acknowledgements}

We kindly thank professor K. Postnov and professor P. Ivanov for unfeigned interest in the subject of this study and for a number of discussions during the preparation of manuscript.
The research is carried out using the equipment of the shared research facilities of HPC computing resources at Lomonosov Moscow State University (see paper \cite{sadovnichy-2013} for its detailed description).
Equipment for the reported study was granted in part by the M. V. Lomonosov Moscow State University Programme of Development.
DNR was supported by grant RSF 14-12-00146 when writing Section \ref{simulations} of this paper. 
VVZ was supported in part by RFBR grant 15-02-08476 and by programme 7 of the Presidium of Russian Academy of Sciences.

\appendix

\section{Representation of cross-rolls}
\label{app_represent}

In order to plot the velocity perturbation in Fig. \ref{fig_4_5}, we take the imaginary part of SFH 
given by equation (\ref{f}), where the Fourier amplitudes of the velocity perturbations, $\hat u_{x,y,z}$, are given by the analytic solution (\ref{u_x_analyt},\ref{u_y_analyt}, \ref{u_z_analyt}).
Explicitly, we have
\begin{equation}
\label{decompose_Im}
\Im \left[ \begin{array}{c}
u_x\\
u_y\\
u_z
\end{array} \right ]=
\left ( \begin{array}{c}
\hat u_x\\
\hat u_y\\
\hat u_z
\end{array} \right ) \sin(\tilde k_x  x+y+k_z z)= {\bf C}_1 + {\bf C}_2 + {\bf C}_3 + {\bf C}_4,
\end{equation}
where 
$${\bf C}_2(\tilde k_x x, y, k_z z) = {\bf C}_1(\tilde k_x x + \pi/2, y - \pi/2, k_z z),$$
$${\bf C}_3(\tilde k_x x, y, k_z z) = {\bf C}_1(\tilde k_x x + \pi/2, y, k_z z - \pi/2),$$
$${\bf C}_4(\tilde k_x x, y, k_z z) = {\bf C}_1(\tilde k_x x, y + \pi/2,k_z z - \pi/2).$$
Therefore, SFH is nothing but a combination of the single velocity pattern, ${\bf C}_1$, with its duplicates translated along the coordinate axes on the one-fourth of corresponding wavelength. 
This velocity pattern has the form 
\begin{equation}
{\bf C}_1(\tilde k_x x,y,k_z z) = 
\left ( \begin{array}{c}
\hat u_x\sin(\tilde k_x  x)\cos(y)\cos(k_z z)\\
\hat u_y\cos(\tilde k_x x)\sin(y)\cos(k_z z)\\
\hat u_z\cos(\tilde k_x x)\cos(y)\sin(k_z z)
\end{array} \right )
\end{equation}
and is represented in Fig. \ref{fig_4_5} by the three mutually perpendicular cross-sections having the common point $\{x=0,\,y=0,\,z=0\}$.

\section{Effect of finite numerical resolution}
\label{app_resolution}

In order to reveal the influence of the numerical viscosity on the transition to turbulence at terminally high $R_{nl}$, we fix $q=-4$ and 
perform a set of additional simulations with exactly the same setup as described in Section \ref{simulations}.
The results are shown in table \ref{tb:subGridEstimation}.

\begin{table}
\caption{The summary of additional numerical simulations for $M = 1$ and $q = -4$.}
\label{tb:subGridEstimation}
\begin{tabular}{ccccc}
$l_z$       & $N$     & $R_{turb}$ & $R_{damp}$    & $R_T$    \\ 
\hline
$1.0$       & $128$   & $370000$   & $310000$      & $340000$ \\

            & $192$   & $120000$   & $100000$      & $110000$ \\

            & $256$   & $120000$   & $100000$      & $110000$ \\
\hline
$2.0$       & $128$   & $17500$    & $15000$       & $16250$  \\

            & $192$   & $21000$    & $17500$       & $19250$  \\
\hline
\end{tabular}
\end{table}

The value of $R_T$ obtained in cubic box substantially drops as we increase resolution up to $N = 192$.
This is to be expected, since at $N=128$ $R_T$ is close to the restriction (\ref{num_Re}).
However, an even higher resolution, $N = 256$, does not lead to further decrease of $R_T$: we conclude that convergence is achieved in this case.
At the same time, $R_T$ in tall box is only weakly affected by the increase of resolution.
Thus, there is a difference between the transition Reynolds numbers in cubic and tall boxes, which cannot be attributed to effect of finite resolution.   


\begin{thebibliography}{60}
\expandafter\ifx\csname natexlab\endcsname\relax\def\natexlab#1{#1}\fi

\bibitem[{{Abramowicz} {et~al.}(1988){Abramowicz}, {Czerny}, {Lasota}, \&
  {Szuszkiewicz}}]{abramowicz-1988}
{Abramowicz}, M.~A., {Czerny}, B., {Lasota}, J.~P., \& {Szuszkiewicz}, E. 1988,
  \apj, 332, 646

\bibitem[{Afshordi {et~al.}(2005)Afshordi, Mukhopadhyay, \&
  Narayan}]{afshordi-2005}
Afshordi, N., Mukhopadhyay, B., \& Narayan, R. 2005, \apj, 629, 373

\bibitem[{{Baggett} \& {Trefethen}(1997)}]{baggett-trefethen-1997}
{Baggett}, J.~S. \& {Trefethen}, L.~N. 1997, Physics of Fluids, 9, 1043

\bibitem[{{Balbus}(2003)}]{balbus-2003}
{Balbus}, S.~A. 2003, \araa, 41, 555

\bibitem[{{Balbus} \& {Hawley}(2006)}]{balbus-2006}
{Balbus}, S.~A. \& {Hawley}, J.~F. 2006, \apj, 652, 1020

\bibitem[{{Belyaev} {et~al.}(2013){Belyaev}, {Rafikov}, \&
  {Stone}}]{rafikov-2013}
{Belyaev}, M.~A., {Rafikov}, R.~R., \& {Stone}, J.~M. 2013, \apj, 770, 67

\bibitem[{{Bisnovatyi-Kogan}(1994)}]{gsbk-1994}
{Bisnovatyi-Kogan}, G.~S. 1994, \mnras, 269, 557

\bibitem[{{Burin} \& {Czarnocki}(2012)}]{burin-2012}
{Burin}, M.~J. \& {Czarnocki}, C.~J. 2012, Journal of Fluid Mechanics, 709, 106

\bibitem[{Butler \& Farrell(1992)}]{butler-farrell-1992}
Butler, K.~M. \& Farrell, B.~F. 1992, Phys. Fluids A, 4, 1637

\bibitem[{{Chagelishvili} {et~al.}(2016){Chagelishvili}, {Hau}, {Khujadze}, \&
  {Oberlack}}]{chagelishvili-2016}
{Chagelishvili}, G., {Hau}, J.-N., {Khujadze}, G., \& {Oberlack}, M. 2016,
  Physical Review Fluids, 1, 043603

\bibitem[{Chagelishvili {et~al.}(2003)Chagelishvili, Zahn, Tevzadze, \&
  Lominadze}]{chagelishvili-2003}
Chagelishvili, G.~D., Zahn, J.-P., Tevzadze, A.~G., \& Lominadze, J.~G. 2003,
  \aap, 402, 401

\bibitem[{{Cherubini} {et~al.}(2010){Cherubini}, {de Palma}, {Robinet}, \&
  {Bottaro}}]{cherubini-2010b}
{Cherubini}, S., {de Palma}, P., {Robinet}, J.-C., \& {Bottaro}, A. 2010, \pre,
  82, 066302

\bibitem[{{Darbyshire} \& {Mullin}(1995)}]{darbyshire-mullin-1995}
{Darbyshire}, A.~G. \& {Mullin}, T. 1995, Journal of Fluid Mechanics, 289, 83

\bibitem[{{Edlund} \& {Ji}(2014)}]{edlund-ji-2014}
{Edlund}, E.~M. \& {Ji}, H. 2014, \prl, 89, 021004

\bibitem[{{Ellingsen} \& {Palm}(1975)}]{ellingsen-palm-1975}
{Ellingsen}, T. \& {Palm}, E. 1975, Physics of Fluids, 18, 487

\bibitem[{{Faisst} \& {Eckhardt}(2004)}]{faisst-eckhardt-2004}
{Faisst}, H. \& {Eckhardt}, B. 2004, Journal of Fluid Mechanics, 504, 343

\bibitem[{{Gogichaishvili} {et~al.}(2017){Gogichaishvili}, {Mamatsashvili},
  {Horton}, {Chagelishvili}, \& {Bodo}}]{mamatsashvili-2017}
{Gogichaishvili}, D., {Mamatsashvili}, G., {Horton}, W., {Chagelishvili}, G.,
  \& {Bodo}, G. 2017, \apj, 845, 70

\bibitem[{{Grossmann}(2000)}]{grossman-2000}
{Grossmann}, S. 2000, Reviews of Modern Physics, 72, 603

\bibitem[{{Grossmann} {et~al.}(2016){Grossmann}, {Lohse}, \&
  {Sun}}]{grossmann-2016}
{Grossmann}, S., {Lohse}, D., \& {Sun}, C. 2016, Annual Review of Fluid
  Mechanics, 48, 150724171740009

\bibitem[{{Hamilton} {et~al.}(1995){Hamilton}, {Kim}, \&
  {Waleffe}}]{waleffe-1995}
{Hamilton}, J.~M., {Kim}, J., \& {Waleffe}, F. 1995, Journal of Fluid
  Mechanics, 287, 317

\bibitem[{{Hawley} {et~al.}(1999){Hawley}, {Balbus}, \&
  {Winters}}]{hawley-1999}
{Hawley}, J.~F., {Balbus}, S.~A., \& {Winters}, W.~F. 1999, \apj, 518, 394

\bibitem[{{Hawley} {et~al.}(1995){Hawley}, {Gammie}, \&
  {Balbus}}]{hawley-gammie-balbus-1995}
{Hawley}, J.~F., {Gammie}, C.~F., \& {Balbus}, S.~A. 1995, \apj, 440, 742

\bibitem[{{Henningson}(1996)}]{henningson-1996}
{Henningson}, D. 1996, Physics of Fluids, 8, 2257

\bibitem[{{Horton} {et~al.}(2010){Horton}, {Kim}, {Chagelishvili}, {Bowman}, \&
  {Lominadze}}]{horton-2010}
{Horton}, W., {Kim}, J.-H., {Chagelishvili}, G.~D., {Bowman}, J.~C., \&
  {Lominadze}, J.~G. 2010, \pre, 81, 066304

\bibitem[{{Inogamov} \& {Sunyaev}(1999)}]{inogamov-sunyaev-1999}
{Inogamov}, N.~A. \& {Sunyaev}, R.~A. 1999, Astronomy Letters, 25, 269

\bibitem[{Ioannou \& Kakouris(2001)}]{ioannou-kakouris-2001}
Ioannou, P.~J. \& Kakouris, A. 2001, \apj, 550, 931

\bibitem[{Johnson \& Gammie(2005)}]{johnson-gammie-2005b}
Johnson, B.~M. \& Gammie, C.~F. 2005, \apj, 635, 149

\bibitem[{Lesur \& Longaretti(2005)}]{lesur-longaretti-2005}
Lesur, G. \& Longaretti, P.-Y. 2005, \aap, 444, 25

\bibitem[{{Lithwick}(2007)}]{lithwick-2007}
{Lithwick}, Y. 2007, \apj, 670, 789

\bibitem[{{Lithwick}(2009)}]{lithwick-2009}
{Lithwick}, Y. 2009, \apj, 693, 85

\bibitem[{{Mamatsashvili} {et~al.}(2016){Mamatsashvili}, {Khujadze},
  {Chagelishvili}, {Dong}, {Jim{\'e}nez}, \& {Foysi}}]{mamatsashvili-2016}
{Mamatsashvili}, G., {Khujadze}, G., {Chagelishvili}, G., {et~al.} 2016, \pre,
  94, 023111

\bibitem[{{Maretzke} {et~al.}(2014){Maretzke}, {Hof}, \&
  {Avila}}]{maretzke-2014}
{Maretzke}, S., {Hof}, B., \& {Avila}, M. 2014, Journal of Fluid Mechanics,
  742, 254

\bibitem[{{Meseguer}(2002)}]{meseguer-2002}
{Meseguer}, {\'A}. 2002, Physics of Fluids, 14, 1655

\bibitem[{Mukhopadhyay {et~al.}(2005)Mukhopadhyay, Afshordi, \&
  Narayan}]{mukhopadhyay-2005}
Mukhopadhyay, B., Afshordi, N., \& Narayan, R. 2005, \apj, 629, 383

\bibitem[{Mukhopadhyay {et~al.}(2006)Mukhopadhyay, Afshordi, \&
  Narayan}]{mukhopadhyay-2006}
Mukhopadhyay, B., Afshordi, N., \& Narayan, R. 2006, Advances in Space
  Research, 38, 2877

\bibitem[{{Narayan} \& {Popham}(1993)}]{narayan-popham-1993}
{Narayan}, R. \& {Popham}, R. 1993, \nat, 362, 820

\bibitem[{{Ostilla-M{\'o}nico} {et~al.}(2014){Ostilla-M{\'o}nico}, {Verzicco},
  {Grossmann}, \& {Lohse}}]{ostilla-2014}
{Ostilla-M{\'o}nico}, R., {Verzicco}, R., {Grossmann}, S., \& {Lohse}, D. 2014,
  Journal of Fluid Mechanics, 748, R3

\bibitem[{{Ostilla-M{\'o}nico} {et~al.}(2016){Ostilla-M{\'o}nico}, {Verzicco},
  \& {Lohse}}]{ostilla-2016}
{Ostilla-M{\'o}nico}, R., {Verzicco}, R., \& {Lohse}, D. 2016, Journal of Fluid
  Mechanics, 799, R1

\bibitem[{{Philippov} {et~al.}(2016){Philippov}, {Rafikov}, \&
  {Stone}}]{philippov-2016}
{Philippov}, A.~A., {Rafikov}, R.~R., \& {Stone}, J.~M. 2016, \apj, 817, 62

\bibitem[{{Popham} {et~al.}(1993){Popham}, {Narayan}, {Hartmann}, \&
  {Kenyon}}]{popham-narayan-1993}
{Popham}, R., {Narayan}, R., {Hartmann}, L., \& {Kenyon}, S. 1993, \apjl, 415,
  L127

\bibitem[{{Popham} \& {Sunyaev}(2001)}]{popham-sunyaev-2001}
{Popham}, R. \& {Sunyaev}, R. 2001, \apj, 547, 355

\bibitem[{{Pringle} \& {Kerswell}(2010)}]{pringle-kerswell-2010}
{Pringle}, C.~C.~T. \& {Kerswell}, R.~R. 2010, Physical Review Letters, 105,
  154502

\bibitem[{Razdoburdin \& Zhuravlev(2015)}]{razdoburdin-zhuravlev-2015}
Razdoburdin, D.~N. \& Zhuravlev, V.~V. 2015, Physics-Uspekhi, 58, 1031

\bibitem[{{Razdoburdin} \& {Zhuravlev}(2017)}]{razdoburdin-zhuravlev-2017}
{Razdoburdin}, D.~N. \& {Zhuravlev}, V.~V. 2017, \mnras, 467, 849

\bibitem[{Reddy \& Henningson(1993)}]{reddy-henningson-1993}
Reddy, S.~C. \& Henningson, D.~S. 1993, Journal of Fluid Mechanics, 252, 209

\bibitem[{Rincon {et~al.}(2007)Rincon, Ogilvie, \& Cossu}]{rincon-2007}
Rincon, F., Ogilvie, G.~I., \& Cossu, C. 2007, \aap, 463, 817

\bibitem[{{Rincon} {et~al.}(2008){Rincon}, {Ogilvie}, {Proctor}, \&
  {Cossu}}]{rincon-2008}
{Rincon}, F., {Ogilvie}, G.~I., {Proctor}, M.~R.~E., \& {Cossu}, C. 2008,
  Astronomische Nachrichten, 329, 750

\bibitem[{Sadovnichy {et~al.}(2013)Sadovnichy, Tikhonravov, Voevodin, \&
  Opanasenko}]{sadovnichy-2013}
Sadovnichy, V., Tikhonravov, A., Voevodin, V., \& Opanasenko, V. 2013, in
  Contemporary High Performance Computing: From Petascale toward Exascale,
  Chapman \& Hall/CRC Computational Science (Boca Raton, United States: Boca
  Raton, United States), 283--307

\bibitem[{{Schartman} {et~al.}(2012){Schartman}, {Ji}, {Burin}, \&
  {Goodman}}]{schartman-2012}
{Schartman}, E., {Ji}, H., {Burin}, M.~J., \& {Goodman}, J. 2012, \aap, 543, 13

\bibitem[{Shakura \& Sunyaev(1973)}]{shakura-sunyaev-1973}
Shakura, N.~I. \& Sunyaev, R.~A. 1973, \aap, 24, 337

\bibitem[{{Shakura} \& {Sunyaev}(1988)}]{shakura-sunyaev-1988}
{Shakura}, N.~I. \& {Sunyaev}, R.~A. 1988, Advances in Space Research, 8, 135

\bibitem[{Shen {et~al.}(2006)Shen, Stone, \& Gardiner}]{shen-2006}
Shen, Y., Stone, J.~M., \& Gardiner, T.~A. 2006, \apj, 653, 513

\bibitem[{{Shi} {et~al.}(2017){Shi}, {Hof}, {Rampp}, \& {Avila}}]{avila-2017}
{Shi}, L., {Hof}, B., {Rampp}, M., \& {Avila}, M. 2017, ArXiv e-prints
  [\eprint[arXiv]{1703.01714}]

\bibitem[{{Stone} \& {Gardiner}(2010)}]{stone-gardiner-2010}
{Stone}, J.~M. \& {Gardiner}, T.~A. 2010, \apjs, 189, 142

\bibitem[{{Stone} {et~al.}(2008){Stone}, {Gardiner}, {Teuben}, {Hawley}, \&
  {Simon}}]{stone-2008}
{Stone}, J.~M., {Gardiner}, T.~A., {Teuben}, P., {Hawley}, J.~F., \& {Simon},
  J.~B. 2008, \apjs, 178, 137

\bibitem[{Trefethen {et~al.}(1993)Trefethen, Trefethen, Reddy, \&
  Driscoll}]{trefethen-1993}
Trefethen, L.~N., Trefethen, A.~E., Reddy, S.~C., \& Driscoll, T.~A. 1993,
  Science, 261, 578

\bibitem[{{Umurhan} \& {Regev}(2004)}]{umurhan-regev-2004}
{Umurhan}, O.~M. \& {Regev}, O. 2004, \apj, 427, 855

\bibitem[{{Waleffe}(1997)}]{waleffe-1997}
{Waleffe}, F. 1997, Physics of Fluids, 9, 883

\bibitem[{Yecko(2004)}]{yecko-2004}
Yecko, P.~A. 2004, \aap, 425, 385

\bibitem[{Zhuravlev \& Razdoburdin(2014)}]{zhuravlev-razdoburdin-2014}
Zhuravlev, V.~V. \& Razdoburdin, D.~N. 2014, \mnras, 442, 870

\end{thebibliography}
\end{document}